\documentclass[iop]{emulateapj}
\usepackage{natbib}
\usepackage{graphicx}

\begin{document} 

\title{An empirical sequence of disk gap opening revealed by rovibrational CO} 

\author{A. Banzatti\altaffilmark{1} and K. M. Pontoppidan\altaffilmark{1}}
\altaffiltext{1}{Space Telescope Science Institute, 3700 San Martin Drive, Baltimore, MD 21218, USA} 

\textit{Accepted by the Astrophysical Journal}

\email{banzatti@stsci.edu}

\begin{abstract} 
The fundamental rovibrational band of CO near 4.7\,$\mu$m is a sensitive tracer of the presence and location of molecular gas in the planet-forming region of protoplanetary disks at 0.01--10\,AU. We present a new analysis of a high-resolution spectral survey (R\,$\sim$\,96,000, or $\sim3.2\,\rm km\,s^{-1}$) of CO rovibrational lines from protoplanetary disks spanning a wide range of stellar masses and of evolutionary properties. We find that the CO emission originates in two distinct velocity components. Line widths of both components correlate strongly with disk inclination, as expected for gas in Keplerian rotation. By measuring the line flux ratios between vibrational transitions $F_{v=2-1}/F_{v=1-0}$, we find that the two velocity components are clearly distinct in excitation. The broad component ($\rm FWHM=50-200\,km\,s^{-1}$) probes the disk region near the magnetospheric accretion radius at $\approx0.05$\,AU, where the gas is hot ($800-1500$\,K). The narrow component ($\rm FWHM=10-50\,km\,s^{-1}$) probes the disk at larger radii of 0.1--10\,AU, where the gas is typically colder (200--700\,K). CO excitation temperatures and orbital radii define an empirical temperature-radius relation as a power law with index $-0.3\pm0.1$ between 0.05--3\,AU. The broad CO component, co-spatial with the observed orbital distribution of hot Jupiters, is rarely detected in transitional and Herbig Ae disks, providing evidence for an early dissipation of the innermost disk. An inversion in the temperature profile beyond 3\,AU is interpreted as a tracer of a regime dominated by UV pumping in largely devoid inner disks, and may be a signature of the last stage before the disk enters the gas-poor debris phase. 
\end{abstract}


\section{INTRODUCTION} \label{sec:intro}
The inner $10$\,AU of protoplanetary disks is thought to be the region in which most exoplanets form \citep{armi}. Initially, the inner disk likely has a smooth radial distribution of gas and dust that is truncated by the stellar magnetic field at the co-rotation radius, within 0.1\,AU. From this point, disk material accretes onto the central star through magnetospheric accretion \citep[e.g.][]{bouvier}. However, over time, inner disks evolve, and gaps and inner holes frequently develop within a few Myr \citep{muzerolle10}. Some of these gaps may be carved by photo-evaporation \citep{alex14}, and others by dynamical clearing by planets coupled with pressure-driven dust migration \citep{zhu11, pinilla15}. The evolutionary window available for planet formation is therefore limited by, and linked to, the time scale of inner disk dispersal of a few Myr \citep{luhman10}. When observing the inner regions of protoplanetary disks and the process of planet formation, a central challenge is that the angular size of these regions is small ($<0.1''$), even for the closest star-forming clouds. Further, the gas temperatures in the planet-forming region typically spans $100-1000$\,K, pushing the main observational line tracers into the near- to far-infrared wavelength range. This leads to the common use of infrared molecular spectroscopy to investigate the properties of inner disk gas.

\begin{figure*}
\includegraphics[width=1\textwidth]{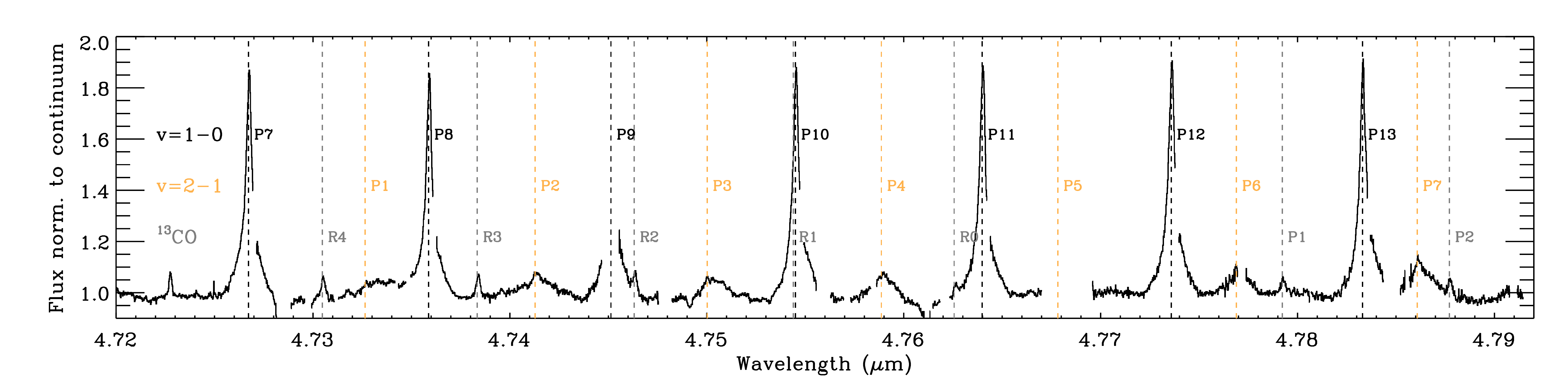} 
\caption{Portion of one of the 4.7\,$\mu$m CRIRES spectra in our sample. Rovibrational P- and R-branch lines of $^{12}$CO and $^{13}$CO are marked in different colors.}
\label{fig: 5um spectrum}
\end{figure*}

The CO molecule is a ubiquitous and sensitive tracer of gas in protoplanetary disks. While pure rotational CO lines observed by millimeter interferometers trace cold gas extending to the outermost disk regions, CO lines from rovibrational transitions at infrared wavelengths are typically excited closer to the central star, where the disk is hotter and exposed to strong infrared (IR) and ultraviolet (UV) radiation fields. Surveys of rovibrational CO emission from the fundamental ($\Delta v=1$) band near 4.7\,$\mu$m have demonstrated that the lines probe the properties of inner disks between the magnetospheric accretion radius and the inner 10 AU \citep[e.g.][]{naji03,blake04,brit07,sal11}. CO line profiles, when spatially- and velocity-resolved, have shown a wealth of CO isotopologues and complex line profiles, as indicative of a range in excitation conditions and of physically distinct emission components \citep{pont11,greg,bast,brown13,vdplas15}. Emission from $^{12}$CO $v$=1-0 and $v$=2-1 lines is strong enough to be detected in the majority of gas-rich disks, demonstrating that CO is often vibrationally excited. In the innermost disk, CO can be efficiently excited by IR photons from the stellar radiation and the local $\mu$m-size hot dust through IR pumping \citep[e.g.][]{Thi13}. Detection of higher vibrational lines in some disks implies vibrational temperatures that are several times larger than the rotational temperatures, providing evidence for the radiative cascade from pumping of directly UV-irradiated CO gas \citep{krotkov}. Detection of rarer isotopologues, including $^{13}$CO that is commonly seen in its $v$=1-0 transitions, has been usually attributed to large column densities of CO gas found in young disks. 

Recent analyses of velocity-resolved CO rovibrational emission in one protoplanetary disk found that the $v$=1-0 lines are a superposition of a broad (FWHM $\sim150$\,km\,s$^{-1}$) and a narrow (FWHM $\sim30$\,km\,s$^{-1}$) components \citep{goto,banz15}. \citet{greg} found evidence for similarly broad and narrow CO components in $\sim10$ more embedded disks, and found that, while the two components are blended in $^{12}$CO $v$=1-0 lines, only the broad component contributes to the $^{12}$CO $v$=2-1 lines, while the $^{13}$CO lines are dominated by the narrow component. 
In this paper, we systematically apply a spectral decomposition analysis to high resolution ($\sim 3\,\rm\,km\,s^{-1}$) spectra of a sample of $\sim40$ protoplanetary disks, to test the hypothesis that CO rovibrational emission in protoplanetary disks may generally be described by two distinct components in velocity and excitation: specifically, a \textit{narrow component}, or NC, and a distinctly \textit{broad component}, or BC. The decomposition allows us to estimate both the location and excitation temperature of the emitting CO gas. While most disks in the sample do show two distinct CO components that get colder at larger disk radii, some show only narrow emission lines and signatures of having cleared-out inner disk regions. Together, this dataset reveals a sequence of inner disk dispersal across a wide range of stellar masses, from 0.3 to 3\,$M_{\odot}$.

\section{OBSERVATIONS} \label{sec:obs}
The disks sample includes 37 disks with 4.7\,$\mu$m spectra obtained with the CRyogenic Infrared Echelle Spectrometer \citep[CRIRES,][]{crires} on the Very Large Telescope (VLT) of the European Southern Observatory (ESO), as part of the ESO Large Program 179.C-0151 \citet{pont_msgr}. Further, one spectrum was added from program 093.C-0432, providing an additional epoch of data for a disk around the variable star EXLup \citep{banz15}. The original survey included disks mostly around young solar-mass stars (M$_{\star}\lesssim1.5$ M$_{\odot}$, T Tauri stars) as well as some around intermediate-mass stars ($1.5<$ M$_{\star}<3$ M$_{\odot}$, mostly Herbig Ae/Be stars). The disks were also selected to cover a range of evolutionary phases, from the early stages when the disks are still embedded in diffuse envelopes to the clearing stages of transitional disks. For this paper, we selected those disks with spectra that show CO lines in emission, indicative of the presence of warm CO gas in the inner part of the disk. The sample includes 30 T Tauri disks and 7 Herbig disks; 2 disks have been observed to have diffuse envelopes, and 7 disks are known to be transitional (see Table \ref{tab: incl}). 

All spectra were reduced using procedures described in \citet{pont11}, and were already presented in several papers, providing general overviews as well as more detailed studies of various subsamples \citep{pont08,pont11,greg,bast,brown13}. The observations were carried out between 2007 and 2010, but consistent instrumental parameters were maintained to ensure homogeneity of the data. The slit width was $\sim0.2"$, providing a spectral resolution of $\sim3.2$\,km\,s$^{-1}$, which is generally confirmed by measured widths of narrow atmospheric absorption lines. Early-type telluric standard stars were observed at similar airmasses to the science targets (typically within 0.1 difference) to allow an accurate telluric correction. The observations were generally obtained using the CRIRES adaptive optics correction. 

The four CRIRES detectors provide a combined instantaneous spectral coverage of $\sim0.1$\,$\mu$m. The observing strategy targeted the $^{12}$CO rovibrational P-branch lines in several non-contiguous settings, with at least one centered between 4.65 and 4.77\,$\mu$m (see Figure \ref{fig: 5um spectrum}). Additionally, a spectral setting covering wavelengths up to 5.0\,$\mu$m was obtained for most targets. The CRIRES spectra do not provide continuous coverage due to gaps from detector chips and the telluric absorption from atmospheric CO lines Doppler-shifted from the targeted lines depending on the month of observation. In some cases, the spectra were repeated by applying a small shift to the central wavelength or taken in different months of the year to compensate for those effects and provide a more complete spectral sampling \citep[see][]{pont11}. This strategy was adopted to obtain complete velocity-resolved CO line profiles.

\begin{figure*}
\includegraphics[width=0.5\textwidth]{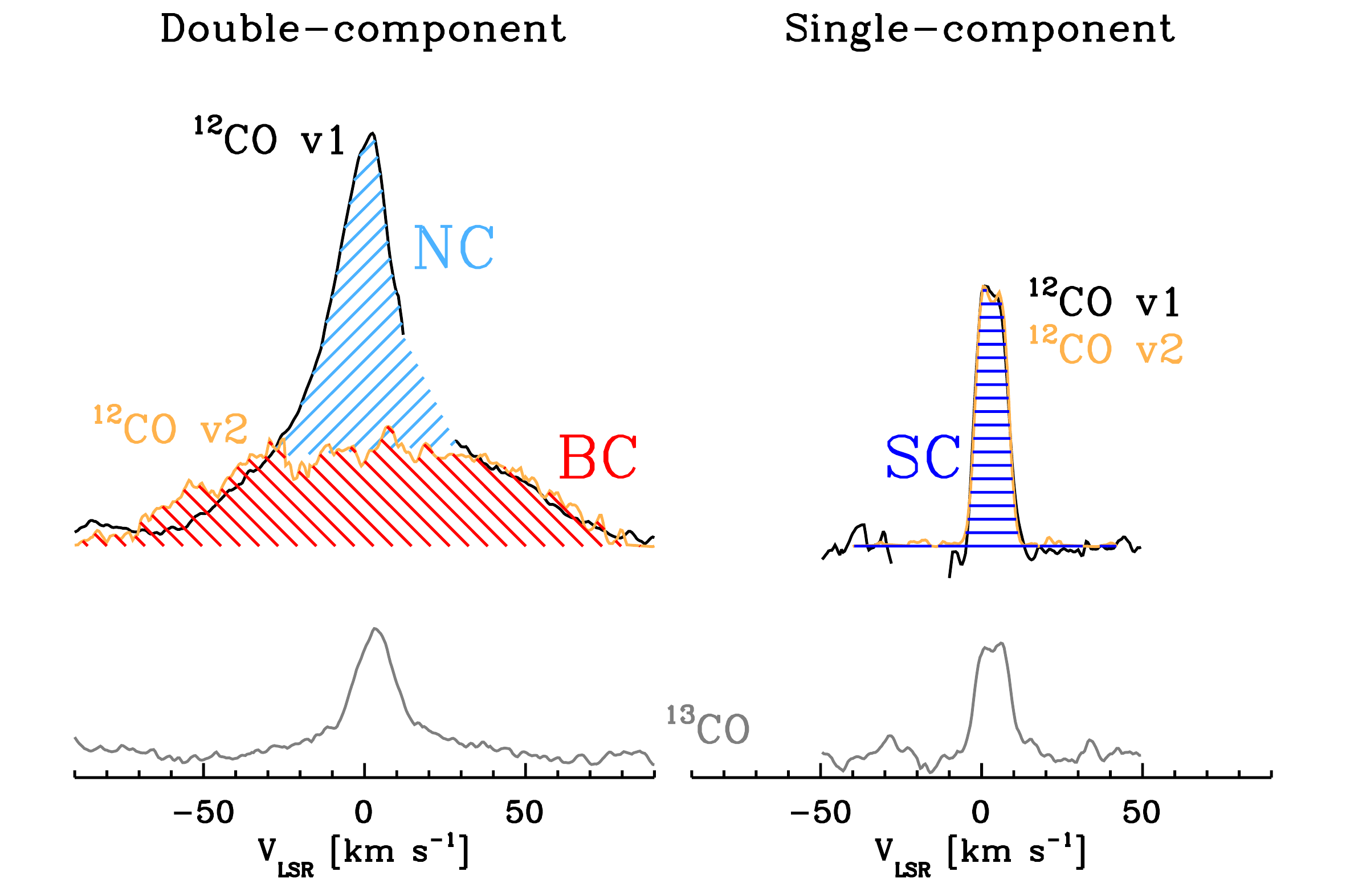} 
\includegraphics[width=0.5\textwidth]{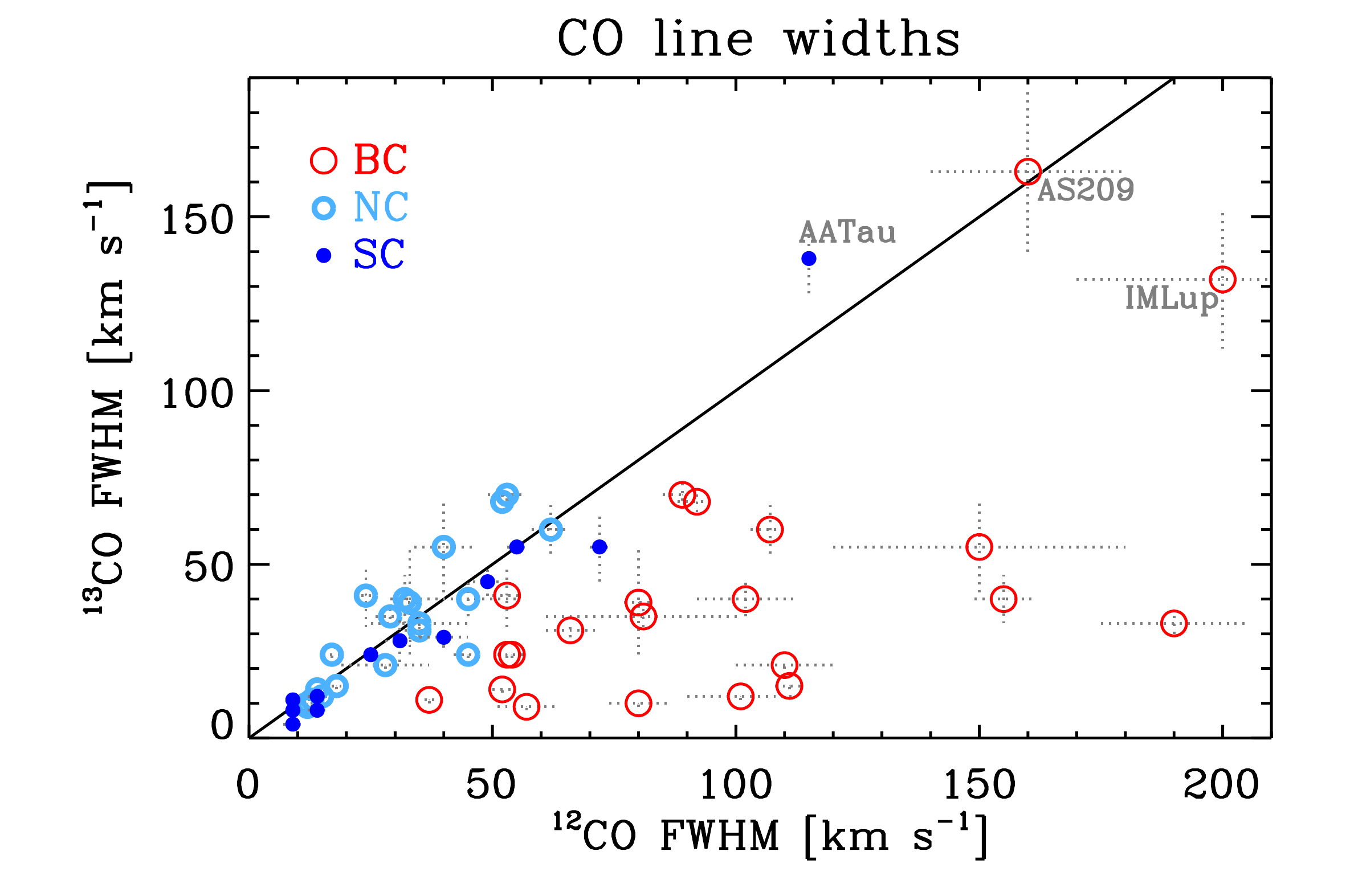} 
\caption{\textit{Left:} Empirical decomposition of CO rovibrational emission from protoplanetary disks into two components, where one (BC) is distinctly broader than the other one (NC). Comparison of $^{12}$CO $v$=1-0 and $v$=2-1 lines shows whether two (double-component disks) rather than one (single-component disks) component is found in a given disk. Decomposition plots for the entire disk sample are provided in Appendix \ref{app: CO_decomp}. \textit{Right:} Comparison between $^{12}$CO and $^{13}$CO line widths. The 1:1 relation is marked with a solid line.}
\label{fig: multi-comp analysis}
\end{figure*}

\section{Analysis of CO spectra} \label{sec:ana}

In this section we describe the methodology applied to the analysis of CRIRES spectra, while comparison of our results to previous work and discussion are provided in Section \ref{sec:disc}. We apply a spectral decomposition algorithm, which essentially consists of the following three basic steps.
\begin{description}
  \item[\textit{Step 1}] High signal-to-noise profiles of $^{12}$CO $v$=1-0 and $v$=2-1 emission are built by stacking multiple lines from different rotational levels.
  \item[\textit{Step 2}] The stacked line profiles are split into two (broad; BC and narrow; NC) components by scaling the $^{12}$CO $v$=2-1 profile to match the wings of the $^{12}$CO $v$=1-0 profile, and then subtracting it. This step reveals if the emission is made of a single (SC) rather than of two components (BC and NC, Figure \ref{fig: multi-comp analysis}).
  \item[\textit{Step 3}] The line widths and the vibrational $F_{v=2-1}/F_{v=1-0}$ ratios are then measured for each detected component.
\end{description}
This procedure is applied to each CO spectrum in our sample. The first two steps are described in Sections \ref{sec:ana1} and \ref{sec:ana2}. The third step provides the data used to find two empirical relations that probe the properties of inner disks (Sections \ref{sec:ana3} and \ref{sec:ana4}). 

\subsection{Stacking of CO line profiles} \label{sec:ana1}
We build high signal-to-noise CO line profiles for each science target by stacking several transitions observed in the 4.7\,$\mu$m spectra. Nominally, the maximum spectral coverage obtained in this survey includes ro-vibrational CO transitions from P(1)--P(32) for $^{12}$CO $v$=1-0, P(1)--P(26) and R(0)--R(6) for $^{12}$CO $v$=2-1, and P(1)--P(23) and R(0)--R(15) for $^{13}$CO $v$=1-0. In practice, the number of transitions available for each disk depends on the spectral coverage obtained for individual targets and on the gaps caused by detector chips and telluric absorption, which in turn depends on the time of year of a given observation. For stacking and averaging line profiles, we use the lines that are most commonly available, which have similar integrated flux, and which are not significantly blended with other transitions. These criteria preserve both the line flux and shape in the stacked profiles, and enforce homogeneity in the analysis method for the entire disk sample. Note that, within the relatively narrow spectral range $4.7-4.9\,\mu$m, lines out of a given vibrational level have very similar fluxes and excitation energies (within 10-30\%) and line profiles depend only very weakly on the rotational quantum number \citep[e.g.][]{sal11,pont11,vdplas15}. For high rotational quantum numbers, this may no longer be true ($J\gtrsim 25$).

Given these constraints, the $^{12}$CO  $v$=1-0 lines that are most suitable for stacking are those between P(7) and P(12), while the $^{12}$CO $v$=2-1 lines are those between P(3) and P(6). As for the $^{13}$CO lines, several can typically be used over the entire spectrum, where not blended with $^{12}$CO. Each line is normalized to its local continuum, re-sampled on a common velocity grid, and the stacked line profiles are built by weighted average of the pixel values from the individual transitions. While the stacked $v$=2-1 line profiles typically have continuous spectral sampling, $v$=1-0 lines often have spectral gaps on one side due to the counterpart telluric absorption lines at the Doppler velocity of the target at the time of observation (see for instance the gap on the red side of the $v$=1-0 line in Figure \ref{fig: multi-comp analysis}). In some disks (CW Tau, DoAr 24E S, HT Lup, VSSG 1), an absorption component is present on top of $^{12}$CO emission in the $v$=1-0 lines. This is typically due to absorption from unrelated foreground gas \citep{greg,pont11,brown13}. In these targets, we exclude pixels affected by the absorption component from the stacked profiles and further analysis.

\begin{deluxetable*}{l l l l c l c l c l}
\tabletypesize{\small}
\tablewidth{0pt}
\tablecaption{\label{tab: incl} Sample properties and disk inclinations.}
\tablehead{\colhead{Name} & \colhead{CO\tablenotemark{a}}  & \colhead{Type\tablenotemark{b}} & \colhead{$M_{\star}$} & Ref & \colhead{$i_{\rm in}$\tablenotemark{c}} & \colhead{Ref} & \colhead{$i_{\rm out}$\tablenotemark{c}} & \colhead{Ref}  & \colhead{Comment} \\
 &   &  & [$M_{\odot}$] &  & [$^{\circ}$] &  & [$^{\circ}$] &   & }
\tablecolumns{10}
\startdata

AATau & S & TT & 0.85 & \textit{m1}  & - & - & 71(1)  & \textit{i1} \\

AS205 N & D & TT & 1.1 & \textit{m2}  & 20(5) & \textit{i2} & 25 & \textit{i3} \\

AS209 & D & TT & 1.4 & \textit{m3}  & - & - & 38--40 & \textit{i3} \\

CrA-IRS2 & D & TT,Em & 1.0 & \textit{m4}  & 17(11) & \textit{this work} & - & - \\

CVCha & D & TT & 2.1 & \textit{m3}  & 30(12) & \textit{this work} & - & - \\

CWTau & D & TT & 1.2 & \textit{m2}  & 28(10) & \textit{this work} & - & - & $i_{\rm{in}}$ agrees with $i_{\rm{jet}} = 41$ from \textit{i4} \\

DFTau & D & TT & 0.5 & \textit{m5}  & 65(15) & \textit{this work} & - & - & $i_{\rm{in}}$ agrees with $i_{\rm{star}} = 60$ from \textit{i5} \\

DoAr24E S & D & TT & 0.7 & \textit{m2}  & 20(5) & \textit{i2} & - & -\\

DoAr44 & S & TT,Tr & 1.3 & \textit{m3}  & - & - & 40(20) & \textit{i3} & \\

DRTau & D & TT & 1.0 & \textit{m2}  & 9(5) & \textit{i2} & 37(3) & \textit{i6} & \\

EC82 & S & TT & 1.0 & \textit{m4}  & 51(18) & \textit{this work} & - & - &  \\

EXLup 08 & D & TT & 0.6 & \textit{m6}  & 44(8) & \textit{this work} & - & - & EX Lup observation from August 2008 \\

EXLup 14 & D & TT & 0.6 & \textit{m6}  & 48(17) & \textit{this work} & - & - & EX Lup observation from April 2014 \\

FNTau & S & TT & 0.35 & \textit{m7}  & - & - & 20(10) & \textit{i7} \\

FZTau & S & TT & 0.7 & \textit{m3}  & 38(15) & \textit{this work} & - & - &  \\

GQLup & D & TT & 0.8 & \textit{m2}  & 65(10) & \textit{i2} & - &  \\

HD135344B & S & H,Tr & 1.6 & \textit{m2}  & 14(4) & \textit{i2} & 23(0.3) & \textit{i8} \\

HD142527 & S & H,Tr & 3.5 & \textit{m2}  & 20(2) & \textit{i2} & 23(8) & \textit{i9} \\

HD144432S & S & H & 1.7 & \textit{m2}  & 27(3) & \textit{i2} & - & - \\

HH100 & D & TT,Em & 0.4 & \textit{m3}  & 26(4) & \textit{this work} & - & - &  \\

HTLup & D & TT & 2.5 & \textit{m3}  & 28(8) & \textit{this work} & - & - &  \\

IMLup & D & TT & 1.0 & \textit{m4}  & - & - & 49(4) & \textit{i10} \\

IRS48 & S & H,Tr & 2.0 & \textit{m3}  & 42(6) & \textit{i11} & 48(8) & \textit{i12} \\

LkHa330 & S & H,Tr & 2.5 & \textit{m2}  & 12(2) & \textit{i2} & 42 & \textit{i13} \\

RNO90 & D & TT & 1.5 & \textit{m2}  & 37(4) & \textit{i2} & - & -\\

RULup & D & TT & 0.7 & \textit{m2}  & 35(5) & \textit{i2} & - & - \\

SCrA S & D & TT & 0.6 & \textit{m3}  & 26(4) & \textit{this work} & - & - \\

SCrA N & D & TT & 1.5 & \textit{m2}  & 10(5) & \textit{i2} & - & -\\ 

SR21 & S & H,Tr & 2.2 & \textit{m2}  & 15(1) & \textit{i2} & 15(0.4) & \textit{i8} & \\

TTau N & D & TT & 2.1 & \textit{m3}  & 20(13) & \textit{this work} & - & - & $i_{\rm in}$ agrees with $i_{\rm disk} < 30$ from \textit{i14} \\

TWCha & S & TT & 1.0 & \textit{m8}  & 67(23) & \textit{this work} & - &- \\

TWHya & S & TT,Tr & 0.7 & \textit{m2}  & 4(1) & \textit{i2} & 12 & \textit{i6} & \\

VSSG1 & D & TT & 0.5 & \textit{m3}  & - & - & 53 & \textit{i3} & \\

VVSer & S & H & 3.0 & \textit{m2} & 65(5) & \textit{i2} & 70(5) & \textit{i15} & \\

VWCha & D & TT & 0.6 & \textit{m3}  & 44(17) & \textit{this work} & - & - \\

VZCha & D & TT & 0.8 & \textit{m3}  & 25(10) & \textit{this work} & - & - \\

WaOph6 & D & TT & 0.9 & \textit{m9}  & - &  & 39 & \textit{i3} \\

WXCha & D & TT & 0.6 & \textit{m8}  & 87(31) & \textit{this work} & - & -

\enddata

\tablecomments{
\tablenotetext{a}{Indication whether two (D) or one (S) component of CO emission are detected.}
\tablenotetext{b}{T Tauri stars (TT) have $M_{\star}\lesssim1.5 M_{\odot}$, and Herbig Ae/Be stars have $M_{\star}>1.5 M_{\odot}$. Disks embedded in diffuse envelopes (Em) or defined as transitional disks by previous work (Tr) are marked accordingly.} 
\tablenotetext{c}{Disk inclinations are reported for both the inner and the outer disk, depending on the technique used. Errors are provided in brackets, where available.}
}

\tablerefs{Stellar masses: (\textit{m1}) \citet{bouv99}, (\textit{m2}) \citet{pont11}, (\textit{m3}) \citet{sal13} and references therein, (\textit{m4}) assumed equal to the median stellar mass in this sample, (\textit{m5}) \citet{beck}, (\textit{m6}) \citet{grasvel}, (\textit{m7}) \citet{mccl13}, (\textit{m8}) \citet{feig}, (\textit{m9}) \citet{and09}. Disk inclinations: (\textit{i1}) \citet{cox} from optical scattered light image, (\textit{i2}) \citet{pont08,pont11} from CO spectro-astrometry, (\textit{i3}) \citet{and09} from resolved mm image, (\textit{i4}) \citet{coffey}, (\textit{i5}) \citet{Unruh}, (\textit{i6}) \citet{isella09} from resolved mm image, (\textit{i7}) \citet{kudo} from NIR scattered light image, (\textit{i8}) \citet{perez14} from resolved mm image, (\textit{i9}) \citet{Avenhaus} from NIR scattered light image, (\textit{i10}) \citet{pinte} from resolved mm and optical scattered light images, (\textit{i11}) \citet{brown12} from resolved NIR CO image, (\textit{i12}) \citet{geers07} from resolved mid-infrared (MIR) image, (\textit{i13}) \citet{brown09} from resolved mm image, (\textit{i14}) \citet{ratzka} from resolved MIR image, (\textit{i15}) \citet{pont07} from resolved MIR image.   }
\end{deluxetable*}

\begin{deluxetable}{l c c c c c c}[ht]
\tabletypesize{\small}
\tablewidth{0pt}
\tablecaption{\label{tab:2co comp} Disks with two CO components.}
\tablehead{  &  \colhead{BC}  &  & &  & \colhead{NC} & \\ 
  \colhead{Name}  & FWHM &  $v2/v1$ &  $R_{\rm CO}$ & $T_{\rm{vib. ex.}}$ & FWHM &  $R_{\rm CO}$ \\
  & [km\,s$^{-1}$] &   &  [AU] & [K] & [km\,s$^{-1}$] &  [AU] }
\tablecolumns{7}
\startdata

AS205 N  & 52 & 0.38 & 0.17 & 970 & 13 & 2.3 \\

AS209  & 160 & 0.91 & 0.07 & 1054 & 33  & 1.7  \\

CrA-IRS2  & 53 & 0.37 & 0.11 & 960 & 17 & 1.1   \\

CVCha  & 102 & 0.25 & 0.18 & 790 & 47  & 0.9   \\

CWTau  & 54 & 0.32 & 0.32 & 890 & 45  & 0.5  \\

DFTau  & 89 & 0.43 & 0.20 & 1060 & 53  & 0.6   \\

DoAr24E S  & 81 & 0.30 & 0.04 & 860 & 29  & 0.4  \\

DRTau  & 37 & 0.39 & 0.06 & 990 & 13  & 0.5   \\

EXLup 08  & 155 & 0.61 & 0.06 & 1420 & 32  & 1.3   \\

 EXLup 14  & 150 & 0.51 & 0.07 & 1220 & 40  & 1.0   \\

 GQLup  & 107 & 0.28 & 0.20 & 830 & 62  & 0.6  \\

 HH100  & 80 & 1.02 & 0.04 & 1200 & 12  & 1.9   \\

 HTLup  & 190 & 0.65 & 0.05 & 1510 & 35  & 1.6 \\

IMLup  & 200 & 1.16 & 0.05 & 1400 & 40  & 1.3 \\

 RNO90 & 92 & 0.21 & 0.23 & 730 & 52  & 0.7  \\

 RULup  & 111 & 0.50 & 0.07 & 1200 & 18  & 2.5 \\

 SCrA S  & 101 & 0.40 & 0.04 & 1000 & 15  & 1.8  \\

 SCrA N  & 57 & 0.30 & 0.05 & 860 & 12  & 1.1 \\

 TTau N & 110 & 0.57 & 0.07 & 1340 & 28 & 1.1 \\

 VSSG1  & 66 & 0.36 & 0.27 & 940 & 35  & 1.0 \\

 VWCha  & 80 & 0.32 & 0.16 & 890 & 33 & 0.9  \\

 VZCha  & 53 & 0.44 & 0.18 & 1080 & 24  & 0.9 \\

 WaOph6  & 133 & 0.33 & 0.07 & 900 & 32  & 1.2 \\

 WXCha  & 140 & 0.45 & 0.11 & 1100 & 67 & 0.5  

\enddata
\tablecomments{$R_{\rm CO}$ is estimated from Kepler's law using the line velocity at HWHM, the stellar mass, and the disk inclination (Table \ref{tab: incl}). $T_{\rm vib. ex.}$ is estimated from $v2/v1=F_{\rm v=2-1}/F_{\rm v=1-0}$ assuming a CO column density of $10^{18}$\,cm$^{-2}$, apart from for AS209, IMLup, and HH100 where $10^{19}$\,cm$^{-2}$ is used (see the Appendix for details). $v2/v1$ and T$_{\rm vib. ex.}$ can be measured for NC only in three disks, where it is detected also in the $^{12}$CO $v$=2-1 lines: AS205 N (0.06, 520 K), DRTau (0.13, 620 K), and SCrA N (0.06, 520 K).}
\end{deluxetable} 

\begin{deluxetable}{l c c c c}[ht]
\tabletypesize{\small}
\tablewidth{0pt}
\tablecaption{\label{tab:1co comp} Disks with a single CO component.}
\tablehead{ \colhead{Name} & \colhead{FWHM} & \colhead{$v2/v1$}  & \colhead{$R_{\rm{CO}}$} & \colhead{$T_{\rm vib. ex.}$} \\
  & [km\,s$^{-1}$] &   &  [AU] & [K] }
\tablecolumns{5}
\startdata

AATau  & 115  & 0.14 & 0.2 & 630  \\

DoAr44  & 72  & 0.11  & 0.4 & 590  \\

EC82  & 49 & 0.03 & 0.9 & 420   \\

FNTau  & 9 & 0.01 & 1.8 & 330  \\

FZTau  & 31 & 0.20 & 1.0 & 720  \\

HD135344B  & 14 & 0.05 & 1.7 & 420  \\

HD142527  & 25 & 0.04 & 2.3 & 460  \\

HD144432S  & 40 & 0.09 & 0.8 & 560  \\

 LkHa330  & 9 & 0.09 & 4.7 & 420  \\

 TWCha  & 65 & 0.16  & 0.7 & 660   \\

 TWHya  & 6 & 0.09 & 0.3 & 500   \\

 VVSer  & 55 & 0.16 & 2.9 & 660 

\\

\hline
\\

 IRS48  & 14 & 0.38 & 16. & 970  \\

 SR21 & 9 & 0.64  & 6.5 & 1480   \\

HD100546 $^{*}$  & 16 & 0.27 & 15. & 810  \\

HD97048 $^{*}$  & 17 & 0.33 & 14. & 900  \\

HD179218 $^{*}$  & 18 & 0.37 & 21. & 950  \\

HD141569 $^{*}$  & 16 & 0.56 & 17. & 1300  \\

HD190073 $^{*}$  & 16 & 0.27 & 6.0 & 810  \\

HD98922 $^{*}$  & 21 & 0.29 & 9.3 & 840  

\enddata
\tablecomments{$R_{\rm CO}$ is estimated from Kepler's law using the line velocity at HWHM, the stellar mass, and the disk inclination (Table \ref{tab: incl}). $T_{\rm vib. ex.}$ is estimated from $v2/v1$ assuming a CO column density of $10^{18}$\,cm$^{-2}$. Disks marked with * are included from \citet{vdplas15}.}
\end{deluxetable}

\subsection{Empirical decomposition into two CO components} \label{sec:ana2}
The presence of separate broad and narrow components is apparent by visual inspection of many disk spectra (e.g. Figure \ref{fig: 5um spectrum}). In practice, the difference is formalized by comparison of the $v$=1-0 and $v$=2-1 stacked line profiles. We scale the  $v$=1-0 to match the broad wings of the $v$=2-1 profile, and then subtract one from the other. The best scaling factor is found by $\chi^2$ minimization between the scaled $v$=1-0 line profile and the $v$=2-1 line profile, excluding the central region of narrow emission, if present. The NC is typically much narrower than BC, such that most of the line profile can be used for the $\chi^2$ minimization. The choice of the wavelength exclusion region is determined iteratively, starting by using the full line profile. If the residuals after subtraction of the two line profiles are consistent with zero over the entire spectral range of the $v$=2-1 line, the NC is not detected and the procedure stops there. In the contrary case, the fit is iterated after exclusion of increasingly large portions at the center of the line profile, until the residuals are composed of a flat continuum and of the NC emission line. Figure \ref{fig: multi-comp analysis} shows a demonstrative sketch of the outcome, while spectral decomposition plots for all disks are shown in Appendix \ref{app: CO_decomp}. 

Two basic parameters are measured from the decomposition procedure: the line width and the vibrational flux ratio $v2/v1 = F_{v=2-1}/F_{v=1-0}$. Line widths are measured at 50\% of the line peak (the full width at half maximum; FWHM) from fits to the line profiles, allowing up to two Gaussian functions to account for double-peaked profiles. The vibrational flux ratio is measured from the scaling factor that matches the $v$=1-0 to the $v$=2-1 lines as explained above, and is exactly equivalent to the integrated line intensity ratio between the $v$=2-1 and the $v$=1-0 lines in each spectrum (further details are provided in the Appendix). It is possible to measure vibrational ratios in all single-component disks and for the BC in all double-component disks. For the NC, $F_{v=2-1}/F_{v=1-0}$ can be measured in only three disks: AS 205 N, DR Tau, and S CrA N. In these disks, the NC can also be identified in the $v$=2-1 lines through fits of multiple Gaussians to the line profile, one for each of the broad and narrow components (plots are shown in the Appendix). The non-detection of the NC in the  $v$=2-1 lines of all other double-component disks is generally consistent with an upper limit on the vibrational flux ratio of $\lesssim 0.2$. Tables \ref{tab:2co comp} and \ref{tab:1co comp} report line widths and vibrational ratios for the entire disk sample.

By implementing this decomposition procedure, we find that all disks fall into two phenomenological classes: the ``single-component" disks (14 of 37), where the $v$=1-0 line profile, once scaled down, is consistent with that of the $v$=2-1 line, and the ``double-component" disks (23 of 37), where the $v$=2-1 profile matches only the wings of the $v$=1-0 profile, revealing an additional narrower component of the $v$=1-0 profile. In the double-component disks, the $^{13}$CO lines are typically significantly narrower than the BC, but match the profile of the NC (Figure \ref{fig: multi-comp analysis}). Only AS 209 and IM Lup have $^{13}$CO lines detected in the BC only, rather than in the NC, while a few disks tentatively show both the BC and NC in $^{13}$CO (CrA-IRS 1, AS 205 N, and RU Lup). In double-component disks, the BC profile shows properties distinct from those of the single-component disks: broader lines ($\approx$ 50--200\,km\,s$^{-1}$), a clear separation in terms of line flux ratios with a boundary at $F_{v=2-1}/F_{v=1-0} \approx 0.2$, and weak or undetected $^{13}$CO lines. Conversely, the NC shares the same line widths ($\approx$ 10--70\,km\,s$^{-1}$), same vibrational ratios $F_{v=2-1}/F_{v=1-0}<0.2$, and same $^{13}$CO detection rate as for CO in 12/14 single-component disks (SC) of our sample.

\subsection{An empirical relation between CO line width and disk inclination} \label{sec:ana3}

\begin{figure}
\includegraphics[width=0.5\textwidth]{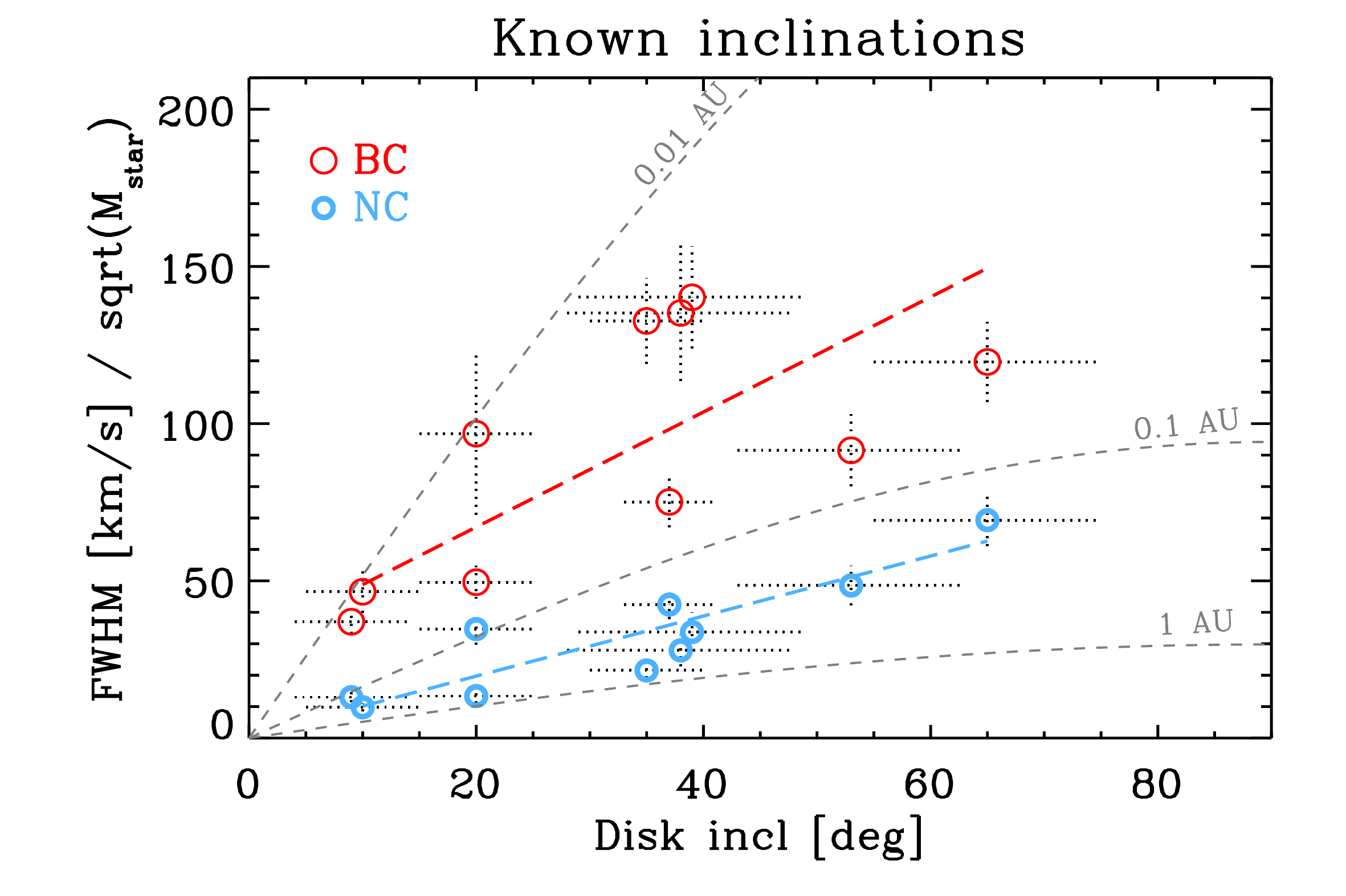} 
\includegraphics[width=0.5\textwidth]{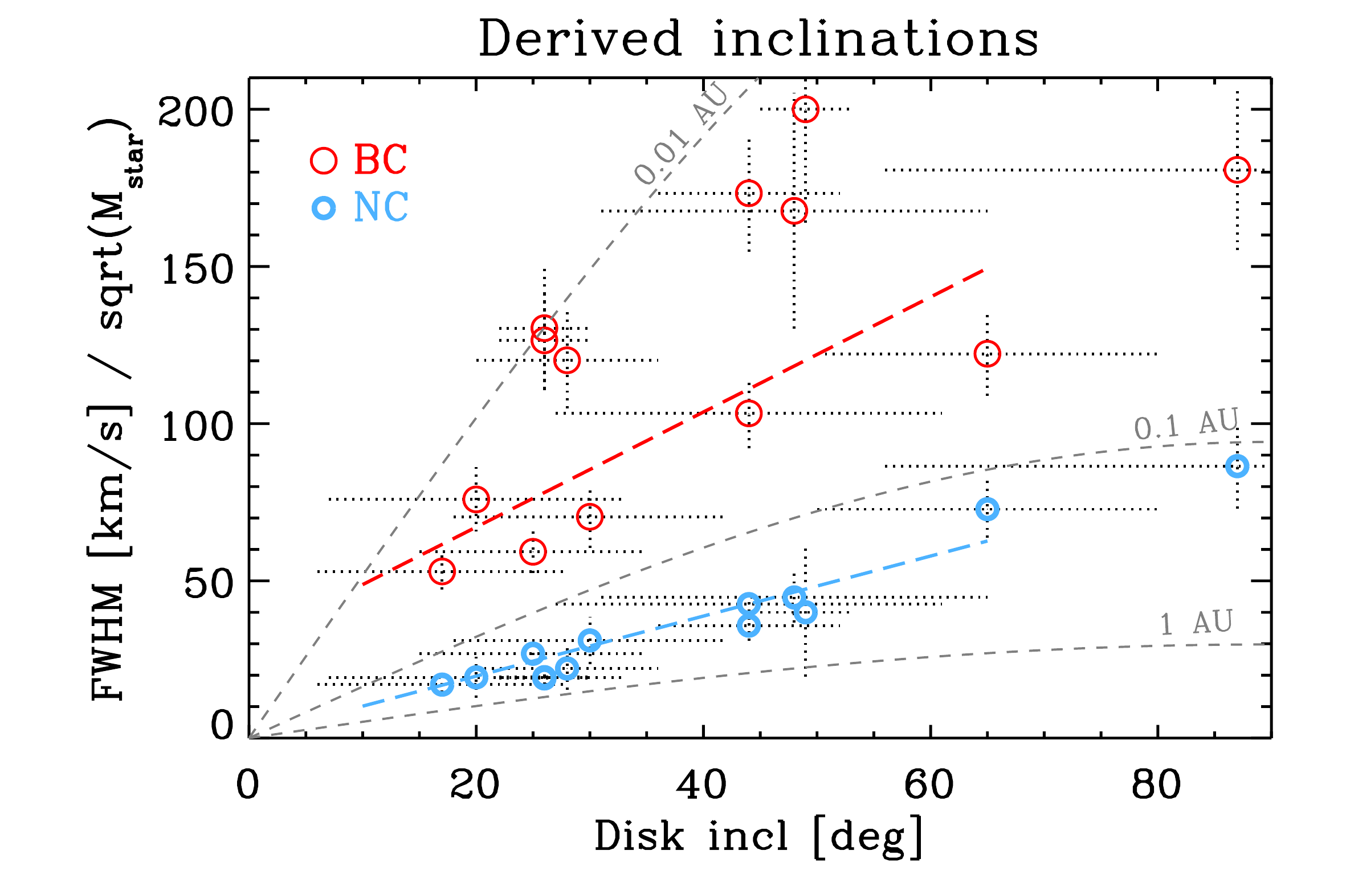} 
\includegraphics[width=0.5\textwidth]{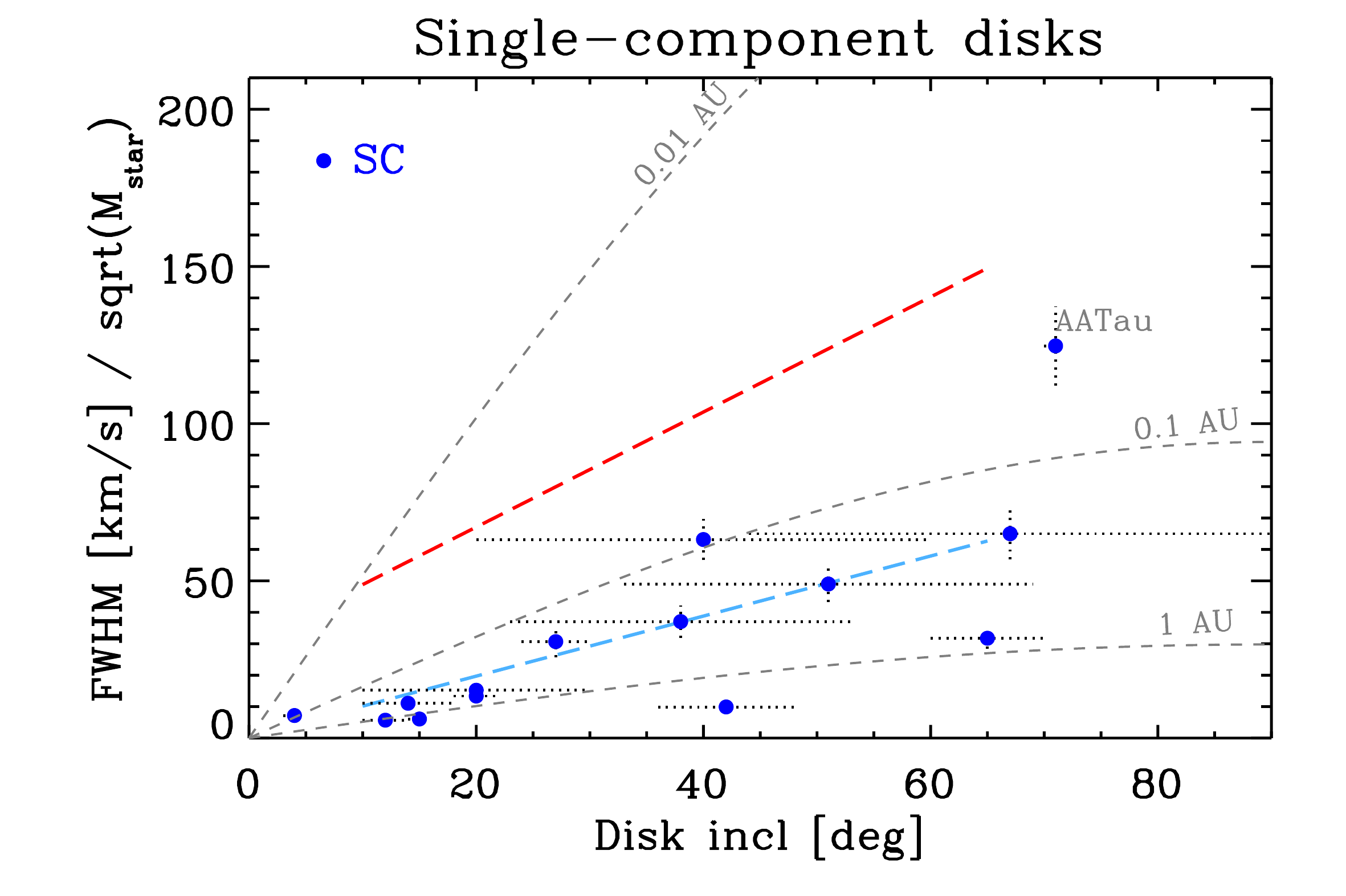} 
\caption{Empirical relations between CO line widths and disk inclinations. \textit{Top}: Linear fits to the distributions of the BC and the NC in disks with known inclinations from in the literature, based on spatially-resolved images. \textit{Middle}: Best-fit disk inclinations derived for disks with no direct inclination measurements from imaging, using the two linear relations from the plot above. \textit{Bottom}: The single-component disks. Grey dashed curves indicate models for gas in Keplerian rotation at three different disk radii around a one solar-mass star, as indicated near each curve.}
\label{fig: fwhm_incl}
\end{figure}

\begin{figure*}
\includegraphics[width=0.5\textwidth]{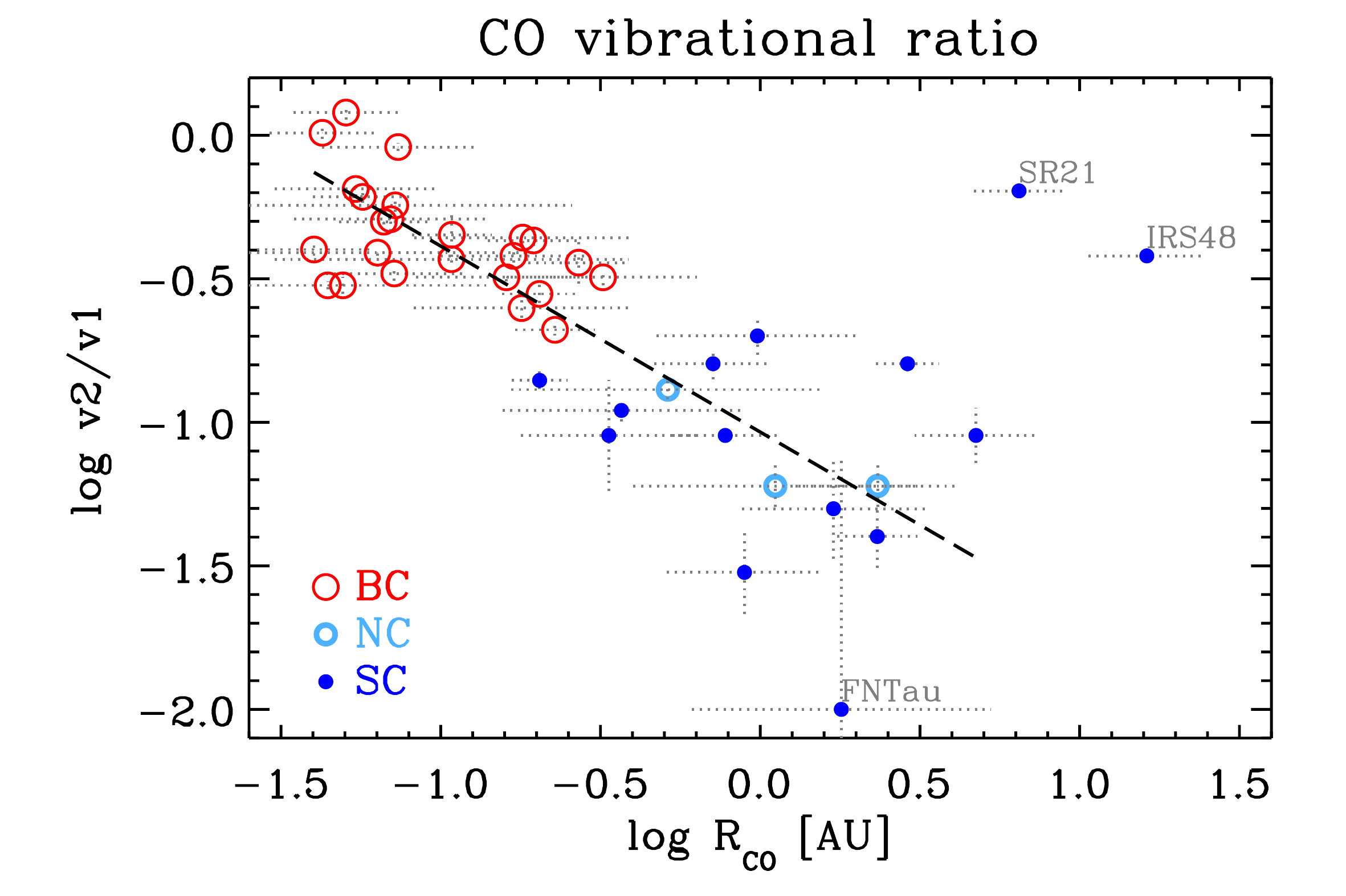} 
\includegraphics[width=0.5\textwidth]{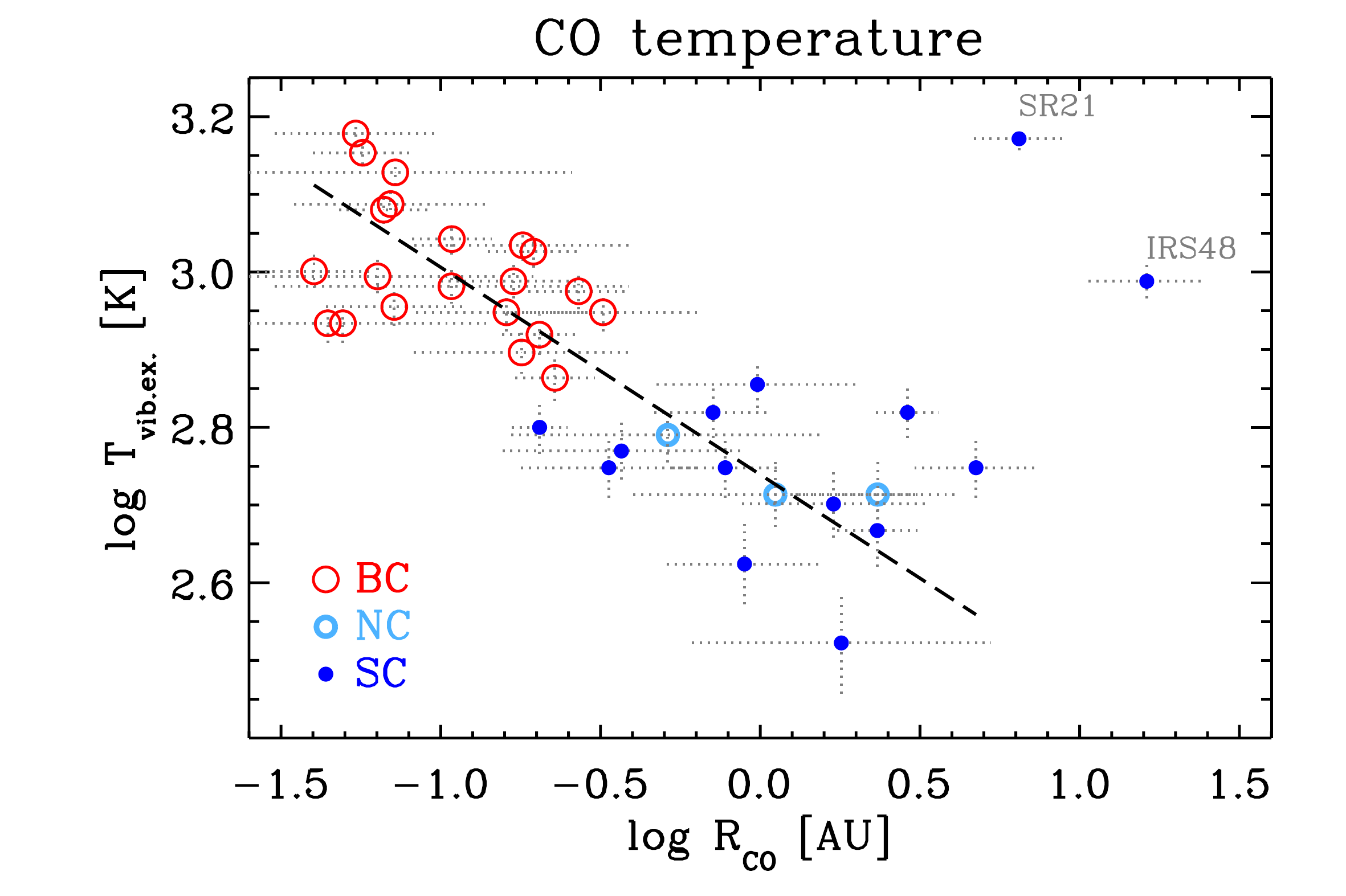} 
\caption{\textit{Left}: Empirical relation found between vibrational flux ratios $F_{v=2-1}/F_{v=1-0}$ and the CO orbital radii $R_{\rm CO}$. The data include all CO components, broad or narrow, detected in the entire disk sample. \textit{Right}: As the left panel, but with the vibrational flux ratio converted into an excitation temperature as explained in the text. This relation is essentially an empirical radial profile for the CO gas temperature in inner disks. SR 21 and IRS 48 are excluded from the fits.}
\label{fig: vibr_rin}
\end{figure*}

When using line widths to infer radial properties of Keplerian disks, it is necessary to correct for inclination effects. Consequently, we collect values of disk inclinations from the literature, where available, for the disks in our sample. We carefully searched for the most reliable estimates obtained using spatially-resolved disk images, either from millimeter interferometry, optical/near-infrared (NIR) scattered-light imaging, or spectro-astrometry from CRIRES. Note that for spectro-astrometry the derived disk inclination is degenerate with an assumed stellar mass. For consistency, we therefore adopt both the stellar masses and the disk inclinations reported by \citet{pont11}. Table \ref{tab: incl} shows the adopted disk inclinations, with notes on the technique used to measure the inclination for each disk. Using ten double-component disks with known inclinations (seven from spectro-astrometry, and three from interferometry), we find strong correlation between the CO line width and disk inclination, with Pearson coefficients of 0.7 for the BC and of 0.9 for the NC. Strong correlation between NIR CO line widths and disk inclination has been reported previously by \citet{blake04} from a sample of five Herbig disks. Correlation between these parameters is expected when the gas is in Keplerian rotation around the star, a condition that is suggested by the high detection rate of double-peaked line profiles in our disk sample ($>70$\%) and that has been proposed since the first surveys of NIR CO emission in disks \citep[e.g.][]{naji03,brit03,blake04}. Note that the two-component decomposition significantly decreases the scatter in this relation relative to what is seen when the line profiles are not fully velocity-resolved as with CRIRES, or if the line profiles were all modeled with a single Gaussian profile to derive average line widths (i.e. without separating BC from NC). 

Gas in Keplerian rotation at a disk radius $R$ produces emission at a maximal velocity $\Delta v$ from the rest frequency of a given CO line following Kepler's law as $\Delta v = \sin (i) \sqrt{GM_{\star}/R}$, where $i$ is the disk inclination, $G$ is the gravitational constant and $M_{\star}$ is the stellar mass. Using this relation, it is possible to derive an emitting radius for the gas in a rotating disk from the observed line widths. In general, emission from higher velocities originate from smaller orbital radii. These velocities, usually taken at the 50\% or 10\% of the line peak value, therefore provide estimates of the disk radii where CO is emitting. We refer to the innermost emitting radius of CO, $R_{\rm in}$, when derived from $1.7\times$HWHM as proposed by \citet{sal11}. In comparison, the velocity at $1.0\times$HWHM provides an estimate of the disk radius $R_{\rm CO}$ that emits the peak line flux \citep[in agreement with the emitting radii derived from spectro-astrometry by][]{pont08,pont11}. Figure \ref{fig: fwhm_incl} shows the measured CO line widths divided by $\sqrt{M_{\star}}$. Removing the stellar mass dependence on the inclination relation leaves only the dependence on the Keplerian radius $R$, which can be compared to a simple model of gas emission in Keplerian rotation around a $1\,M_{\odot}$ star (shown in Figure \ref{fig: fwhm_incl} at three different disk radii). 

In these terms, the BC has $R_{\rm{in}}$ between 0.01 and 0.1 AU. This is consistent with the disk region of the corotation and magnetospheric accretion radii for solar-mass stars \citep[e.g.][]{bouvier}, as found by previous studies of rovibrational CO emission from disks \citep[e.g.][]{naji03,sal11}. The NC, instead, emits from larger disk radii, with $R_{\rm{in}}$ between 0.1 and 1 AU. The single-component disks show CO emission at radii similar to, or larger than, those of the NC.

In Figure \ref{fig: fwhm_incl} we show linear fits to the BC and NC in disks with known inclinations, obtained using the Bayesian method by \citet{kelly}. The regression takes into account the uncertainties on both variables, as well as the intrinsic scatter in the regression relation due to physical quantities not explicitly accounted for. Typical errors on the line widths are $\lesssim 7\rm \,km\,s^{-1}$ for the BC and $\lesssim 4\rm \,km\,s^{-1}$ for the NC, the stellar masses are assumed to be known to 20\% and the disk inclinations are typically known to accuracies better than 10\%. The best-fit parameters and their errors are taken as the median and standard deviation of the posterior distributions of the regression parameters. The median intercept and slope are used to plot the lines in Figure \ref{fig: fwhm_incl}, while the posterior distributions are plotted in the Appendix. 

The single linear fits shown in Figure \ref{fig: fwhm_incl} assume that differences in $R_{\rm{in}}$ due to the dependence on stellar and disk luminosity for either the BC and the NC is weak relative to the inclination dependence. This is supported by the Bayesian regression of line widths versus disk inclinations, which confirms strong correlation ($> 0.8$) for both the BC and NC, although the uncertainties in the regression parameters for BC are 3--4 times larger than for NC. This is also consistent with previous results for disks around solar-mass stars where $R_{\rm{in}}$ was found to vary only within a factor of a few \citep{sal11}, and with the theoretical expectation that the inner radius increases only with $\sqrt{L_{\star}}$. The residual intrinsic scatter on the linear relation for the BC is four times larger than for the NC, which may point to a remaining dependence on the stellar and (variable) accretion luminosity.

\newpage

\subsection{New estimates of disks inclinations}
Given the empirical line width-inclination relation, it is possible to estimate inclinations for disks with high resolution CO spectroscopy even when resolved imaging is not yet available. In doing so, it is assumed that the two relations found for the BC and NC represent a general property of rovibrational CO emission in the double-component disks. The advantage of having two relations, one for each velocity component, is that for a given double-component disk there are two constraints to derive one parameter. We take the measured values of the BC and NC line widths in each disk and use them to derive the disk inclination by using the regression coefficients $a,b$ from the empirical linear relations:

\begin{equation}
i = ( {\rm FWHM} / \sqrt{M_{\star}} - a ) / b,
\end{equation}

where $a=31\pm30$ and $b=1.8\pm0.9$ for the BC, and $a=0.6\pm8.8$ and $b=1.0\pm0.3$ for the NC. We derive two inclination values, one for the BC and one for the NC, and adopt the mean of the 1-$\sigma$ range in common between the two. Among the single-component disks, only three have unknown inclination: EC 82, FZ Tau, TW Cha. For these disks, we adopt the NC relation, since single-component disks with known inclinations span disk radii similar to those of the NC of double-component disks (Figure \ref{fig: fwhm_incl}). Following this procedure, we estimate an inclination for a total of fifteen new disks. These values are included in Table \ref{tab: incl}.

\subsection{A diagram of CO vibrational ratios and orbital radii} \label{sec:ana4}
Using the inclinations described in the previous section, we derive the CO emitting radii for all disks in the sample. In Figure \ref{fig: vibr_rin} we show a diagram made with the vibrational line flux ratios plotted against the CO emitting radii. The diagram at the left of the Figure shows a strong logarithmic anti-correlation between $R_{\rm{CO}}$ and $F_{v=2-1}/F_{v=1-0}$, with a Pearson coefficient of -0.9. In addition to spanning spatially separated $R_{\rm CO}$ ranges, as already shown in Figure \ref{fig: fwhm_incl}, the BC and NC clearly span distinct values in $F_{v=2-1}/F_{v=1-0}$. Here too, the single-component disks have similar values to those of the NC. Two obvious outliers are SR 21 and IRS 48, both transitional disks with large inner gaps, which have vibrational excitation temperatures as high as the BC at much smaller disk radii (they will be discussed in Section \ref{sec:disc}). Excluding them, Figure \ref{fig: vibr_rin} demonstrates that the vibrational flux ratio decreases with disk radius as:

\begin{equation} 
\frac{F_{v=2-1}}{F_{v=1-0}} \propto R^{-0.7 \pm 0.1}. 
\end{equation}

The vibrational line flux ratios measure how much the vibrational level $v$=2 is populated as compared to $v$=1, which in turn is a measure of the vibrational excitation temperature $T_{\rm vib. ex.}$ \citep[e.g.][]{Thi13}. The relation between $F_{v=2-1}/F_{v=1-0}$ and $R_{\rm CO}$ can thus be converted into a radial profile for the CO vibrational excitation temperature in inner disks. The conversion of the vibrational flux ratio to a temperature depends on the CO column density, so the model-independent relation is that using the vibrational flux ratio (Eqn. 2). To estimate the corresponding temperature relation, we use a model of a slab of gas in thermal equilibrium, where the level populations are defined by a single temperature following the Boltzmann distribution. Other analyses of CO rovibrational emission from disks consistently find a relatively narrow range of column densities. Consequently, we adopt a single $^{12}$CO column density of $10^{18}$\,cm$^{-2}$, as an order-of-magnitude value representative of the range found in previous studies \citep[$10^{17}$--$10^{19}$\,cm$^{-2}$,][]{blake04,greg,sal11,banz15,vdplas15}. The details of the conversion are described in the Appendix \ref{app: vibr_convers}. The resulting temperature radial profile is:

\begin{equation}
T_{\rm vib. ex.} \propto R^{-0.3 \pm 0.1}
\end{equation}

If disks without previously known inclinations are excluded, the exponent becomes $-0.2\pm 0.1$, within the error, indicating that the result is not biased by our empirical derivation of disk inclinations. The excitation temperatures estimated for each disk are reported in Tables \ref{tab:2co comp} and \ref{tab:1co comp}. The nominal uncertainty on the temperatures, once the column density is fixed, is less than 50 K as due to the uncertainty on the vibrational ratios (which is typically $<10\%$).

\citet{greg} proposed that the non-detection of $^{13}$CO lines in the BC, combined with the curvature seen in rotation diagrams, constrained the BC column density to within $10^{17.5}$--$10^{18}$\,cm$^{-2}$. On the other hand, rotation diagrams of $^{13}$CO lines were found to require column densities of $10^{18.5}$--$10^{19}$\,cm$^{-2}$ for the NC. If we relax our assumption and allow the BC to have columns as low as $10^{17}$\,cm$^{-2}$, and NC as large as $10^{19}$\,cm$^{-2}$, then the temperatures will decrease/increase for the NC/BC, respectively, and the overall temperature profile may exhibit a steeper index up to $q\approx0.5$ (see Appendix \ref{app: vibr_convers}).

\section{Discussion} \label{sec:disc}
\subsection{Origin of the broad and narrow CO components in disks}
The vibrational excitation temperatures of the BC match those previously found from rotational analyses of $^{12}$CO lines \citep[700--1700\,K, e.g.][]{sal11}, while the CO in the NC and SC disks typically have lower temperatures, corresponding to those determined for $^{13}$CO line fluxes \citep[200--600\,K, e.g.][]{brown13}. This is consistent with our finding that the $^{13}$CO line profiles match those of the NC, as indicative of probing a similar disk region. In turn, this suggests that the NC is formed in an optically thick layer, leading to high $F(^{13}{\rm CO})/F(^{12}{\rm CO})$ ratios, while the BC must be more optically thin. A similar conclusion was reached for disks in embedded young stars \citep{greg}. 

The radii derived from the BC span the range found previously for disks around solar-mass stars (0.01--0.1\,AU) while the NC and SC span the range found for Herbig and transitional disks (0.1--10 AU). This is consistent with the finding that double-component disks (where BC is detected) are primarily disks around low-mass stars, while the single-component disks are primarily Herbig and transitional disks. The NC and part of SC share similar properties in terms of line widths, orbital radii, and vibrational ratios (Section \ref{sec:ana} and Table \ref{tab:co summary}), and they may trace the same gas component, regardless of the presence of a BC.

\begin{deluxetable}{l c c c c c}
\tabletypesize{\small}
\tablewidth{0pt}
\tablecaption{\label{tab:co summary} Summary of rovibrational CO emission properties.}
\tablehead{ \colhead{CO comp.} & \colhead{FWHM} & \colhead{$v2/v1$}  & \colhead{R$_{\rm{co}}$} & \colhead{T$_{\rm vib. ex. }$} & \colhead{$^{13}$CO} \\
  & [km\,s$^{-1}$] &   &  [AU] & [K] & }
\tablecolumns{6}
\startdata
Broad (BC)  & 50--200  & 0.2--0.6 & 0.04--0.3 & 800--1500 & 3/23 \\
Narrow (NC)  & 10--70  & $<0.2$  & 0.2--3 & $<700$ & 18/23 \\
Single I (SC) & 6--70 & 0.01--0.2 & 0.2--5 & 300-700 & 9/12 \\
Single II (SC) & 9--20 & 0.3--0.6 & 6--20 & 800-1500 & 8/8 
\enddata
\tablecomments{Single-component disks are separated into two groups based on their vibrational ratios, with $F_{v=2-1}/F_{v=1-0} < 0.2$ found for the IR regime and $0.3 < F_{v=2-1}/F_{v=1-0} < 0.6$ found for the UV regime shown in Figure \ref{fig: tempprofs_compar}, including the 6 Herbig disks from \citet{vdplas15}. The last column reports $^{13}$CO detection fractions. }
\end{deluxetable} 

While a disk origin for the BC is supported by the common double-peak profiles, as well as by a Keplerian signal seen in spectro-astrometry \citep{pont11}, the origin of the NC is less clear. A disk wind has been proposed for single-peak $^{12}$CO $v$=1-0 line profiles (those dominated by the NC, e.g. AS205 N) to reproduce the sub-Keplerian motions of the line peak, observed with spectro-astrometry \citep{bast,pont11}. In one case, a corresponding large-scale disk wind has possibly been seen with ALMA \citep{sal14}. An observational characteristic of a disk wind launched at small radii, and preserving angular momentum as it expands, is that it the gas contributing to the line becomes increasingly sub-Keplerian closer to the line center. Specifically, the line wings are exactly Keplerian, while the inner few km\,s$^{-1}$ are strongly sub-Keplerian, tracing radii significantly smaller than what their velocity would suggest. This could explain why observables constructed from the line wings of both the broad and narrow components, as we do here, yield results consistent with a nearly Keplerian disk.

Broad and narrow components that may be similar to those of the rovibrational CO lines have been found in the 6300\AA [OI] line by \citet{rigl13}. These authors suggest a disk origin for the broad component, and an unbound cool disk wind as the origin of the narrow component. It is not clear whether the narrow components of [OI] and CO are generally tracing the same disk wind, but they share close resemblance in at least a few disks. For the disks in common between this work and \citet{rigl13}, the FWHM of NC in CO and [OI] agrees to within $<3$\,km\,s$^{-1}$ in DR Tau, VZ Cha, RU Lup, TW Hya, and CW Tau; in DR Tau, the same is true also for the BC of CO and [OI].

\subsection{The radial temperature profile as a tracer of excitation mechanisms} \label{sec: disc2}
\begin{figure}
\includegraphics[width=0.5\textwidth]{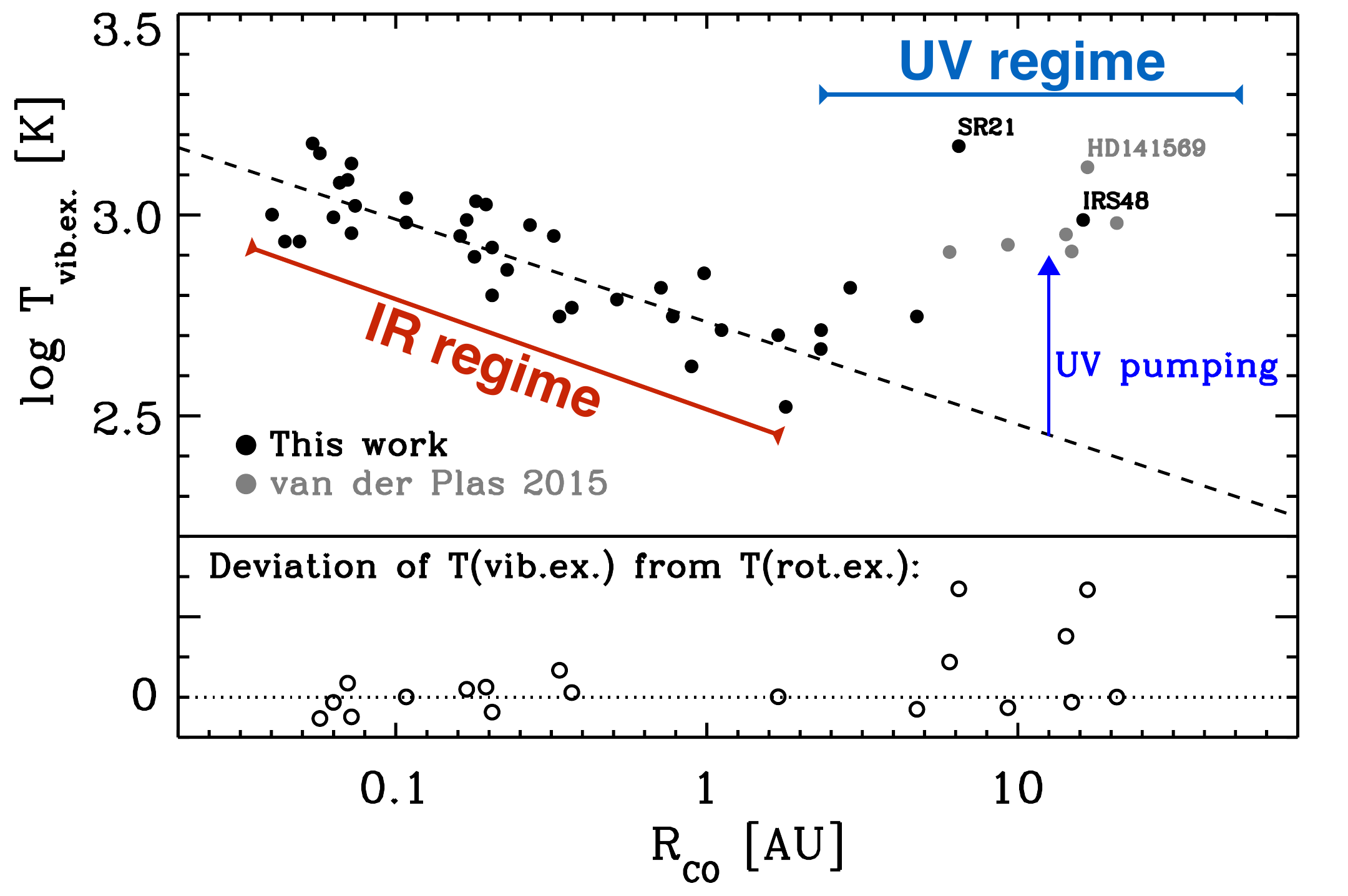} 
\caption{CO temperature-radius diagram constructed by combination of CO rovibrational spectra from this work and spectra from the survey by \citet{vdplas15}. The datapoints from this work are the same as shown in Figure \ref{fig: vibr_rin}.} 
\label{fig: tempprofs_compar}
\end{figure}

In Section \ref{sec:ana4} we found a strong logarithmic anti-correlation between the CO vibrational excitation temperature and emitting radius. This relation corresponds to an empirical temperature profile of a typical inner disk surface. However, it was noted that two disks from the sample with large CO emitting radii, the transitional disks SR 21 and IRS 48, clearly depart from the relation, showing significantly higher vibrational ratios at their respective CO emitting radii than the relation predicts. To explore this further, we add 6 disks from the CRIRES survey of 12 disks around intermediate-mass ($M_{\star}\approx 2-3\,M_{\odot}$) stars from \citet{vdplas15}. This survey presents 4.7\,$\mu$m spectroscopy with a similar experimental design and wavelength coverage as our survey. Specifically, we include the full subset of disks for which lines are detected from both $^{12}$CO $v$=1-0 and $v$=2-1 transitions in a wavelength range similar to that of our survey. Following the same procedure as for other disks, we derive $R_{\rm CO}$ and $T_{\rm vib. ex.}$ for the disks around HD100546, HD179218, HD190073, HD97048, HD141569, HD98922 (included in Table \ref{tab:1co comp}). Their locations in the temperature-radius diagram are included in Figure \ref{fig: tempprofs_compar}, which shows that disks around intermediate-mass stars have vibrational temperatures that are typically higher at a given radius than the relation defined by the lower-mass stars at smaller disk radii. It is clear that SR 21 and IRS 48 belong to this vibrationally hot class of disks. The combination of CRIRES surveys of disks spanning a wide range of stellar masses therefore demonstrates the existence of two trends: a vibrational temperature decrease out to radii of $\approx3$\,AU followed by a temperature increase at radii $>3$\,AU. We note that the remaining six disks from van der Plas et al. (2015) lack $v$=2-1 measurements due either to poor telluric corrections and low signal-to-noise, or the line widths and disk inclinations put them in the IR regime of our diagram ($<3$\,AU), where low $v$=2-1 line fluxes are expected.

\begin{figure*}[ht]
\centering
\includegraphics[width=1\textwidth]{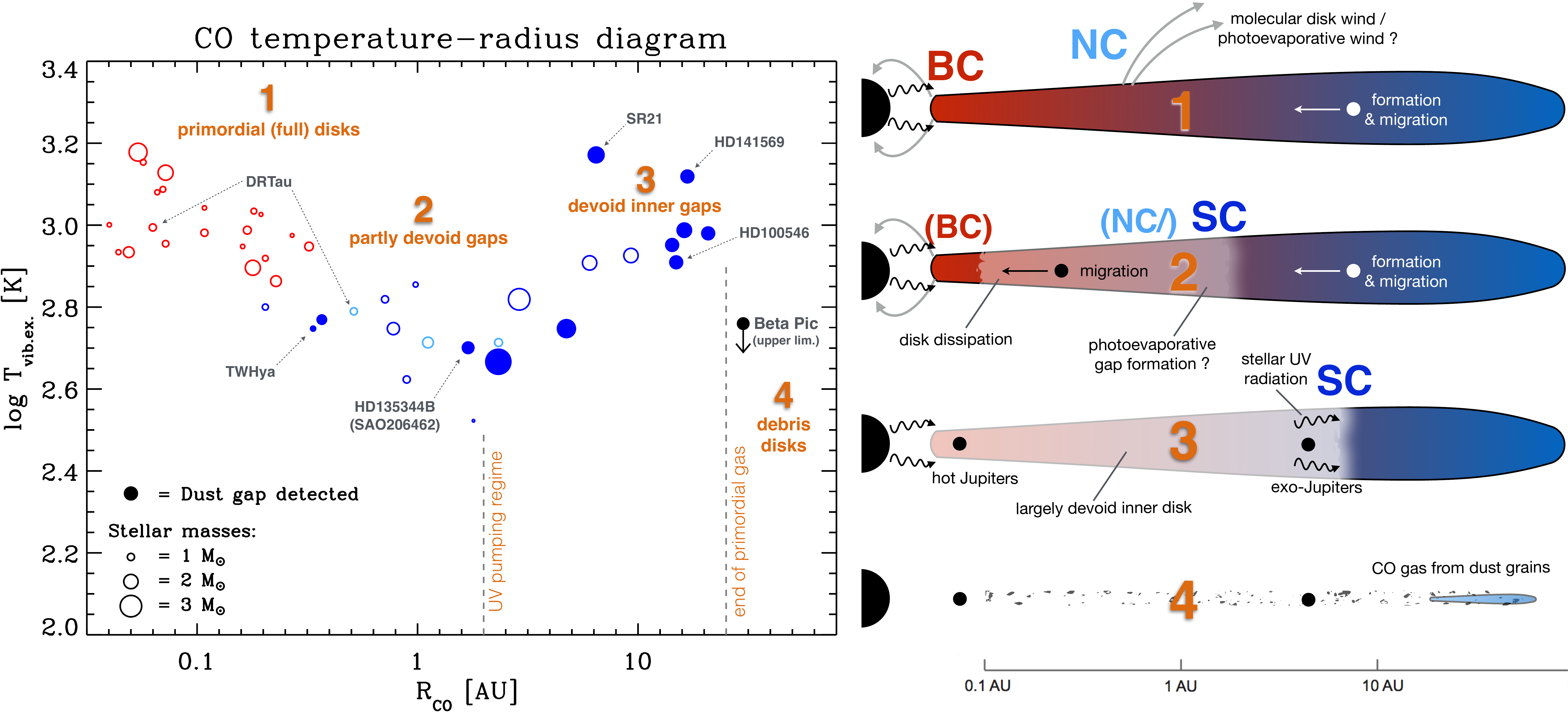} 
\caption{\textit{Left}: The CO temperature-radius diagram is illustrated in the context of disk dispersal and planet formation as discussed in Section \ref{sec:disc}. The data points are identical to those shown in Figure \ref{fig: tempprofs_compar}, but the stellar masses are indicated by the symbol size. Disks with previous detection of gaps in the dust emission are marked with filled symbols. An upper limit on the vibrational temperature for rovibrational CO detected in the debris disk around Beta Pic is also included \citep{trout}. \textit{Right}: A sketch illustrating the evolution of a protoplanetary disk \citep[based in part on][]{alex14}, matched to the different regions of the CO temperature-radius diagram.} 
\label{fig: final}
\end{figure*}

Analyses by \citet{brit03}, \citet{blake04}, and \citet{vdplas15} included two possible excitation regimes for CO rovibrational emission: in the inner disk, IR pumping due to the local dust continuum radiation, and at larger disk radii, where the dust temperature is too low to generate sufficient amounts of 4.7\,$\mu$m photons, pumping by UV photons from the star and/or accretion shock. The specificity of UV pumping is to populate very high vibrational levels ($v\ge 6$) by decay of an excited electronic state \citep{krotkov}, while IR pumping drives the CO level population to equilibrate to the local color temperature of the dust. At dust temperatures of 500-1000\,K, this will only populate the $v$=1 and $v$=2 levels. Indeed, Figure \ref{fig: tempprofs_compar} can be interpreted as evidence for a transition between the IR and UV pumping regimes near 3\,AU. The ``IR regime", at small disk radii, and the ``UV regime", at larger disk radii, are correspondingly marked in the Figure. Further evidence is provided by comparison to rotational temperatures T$_{\rm rot. ex.}$ derived in the same disks from previous studies \citep{greg,sal09,sal11}. Figure \ref{fig: tempprofs_compar} shows that $T_{\rm vib. ex.}\simeq T_{\rm rot. ex.}$ in the IR regime, indicating that the CO rovibrational population is thermalized to the local dust temperature in the inner disk \citep{blake04}. Conversely, in the UV regime, $T_{\rm rot. ex.} < T_{\rm vib. ex.}$ \citep{Thi13}. This is consistent with the finding of others, for instance \citet{brit03} found $T_{\rm rot. ex.} = 200$\,K and $T_{\rm vib. ex.} > 2000$\,K in the disk of HD141569. Note that, while the vibrational temperatures of the $v$=2-1 transitions are higher than the rotational temperature in the UV regime, they are typically much lower than the vibrational temperature of even higher-lying levels ($v$=6-5 can be as high as 6000\,K). Therefore, while we use the $v$=2-1 temperature since it is easy to measure in most disks, our results are consistent with previous work citing very high temperatures of high-lying vibrational levels. 
\\

\subsection{An empirical sequence of disk dispersal} \label{sec: disc4}
In Figure \ref{fig: final} we illustrate how the temperature-radius diagram of rovibrational CO emission resembles key aspects of the current theoretical picture of protoplanetary disk dispersal. In the phenomenological class of the double-component disks, the BC is observed from a hot disk region extending to the magnetospheric accretion radius at $\approx0.05$\,AU \citep{bouvier}, demonstrating the presence of abundant gas at the smallest radii dynamically permissible for Keplerian orbits. These disks are likely to be in a primordial phase, before giant planets have fully formed (sketch 1). DR Tau is an example of a primordial disk as based on its CO emission. 

The next stage in the CO sequence leads to the suppression and eventual disappearence of BC (sketch 2), strongly suggesting the opening of partly devoid inner gaps or an inner region depleted in gas compared to the primordial disks. In fact, it has been observed by monitoring CO spectra of a strong episodic accretor, EX Lup, that the BC gets significantly weaker relative to NC after an accretion outburst that depleted the gas mass in the inner disk by an order of magnitude \citep[][]{banz15}. In this stage, double- and single-component disks overlap in the diagram. NC is detected from fractions of an AU up to $\approx3$\,AU. At similar radii, and even larger, CO is observed in single-component disks, where the BC is not detected. TW Hya is an example of this second class of disks, which probably has significant overlap with the ``pre-transitional'' disks as defined from dust emission \citep{espaillat07}. Indeed, for TW Hya, an inner gap in the optically thick dust disk has been detected \citep{calvet}. Note that a non-detection of the BC does not directly imply the complete absence of dust or gas at $\lesssim$0.1 AU. Residual gas and dust in the innermost region, which absorbs most of the direct UV radiation from the star, may likely explain why the temperature profile in some single-component disks continues along an IR-pumped power law. Photo-evaporation probably plays a role in opening disk gaps, and it is interesting to note that UV photo-evaporation is expected to take over at a few AU \citep{alex06,alex14}, close to where the transition between the IR and UV regimes is found in the CO diagram. Other clearing mechanisms may dominate the disk dissipation at smaller disk radii, including disk accretion and dust evolution to growing planetesimals and protoplanets \citep[e.g.][]{gorti15}.

The transition to a UV-dominated regime of CO rovibrational excitation defines a third class of disks, as discussed in Section \ref{sec: disc2}. In order for UV radiation to excite the gas to high vibrational temperatures beyond a few AU, the inner disk region must be largely devoid of shielding dust. We identify these disks as having large inner gaps, with very little amounts of residual material in the inner disk (sketch 3). This interpretation is consistent with that proposed by \citet{vdplas15}, who also suggested that UV-pumped CO emission may be a distinct signature of devoid inner gaps, and \citet{maask13}, who imaged large inner dust gaps in two Herbig disks in the UV branch of our diagram, one of which was previously unknown (in HD97048). In total, large inner gaps in the dusty disks have been detected in up to 6 out of 8 disks in the UV branch so far, and for the remaining two (HD98922 and HD190073) the current lack of resolved imaging may simply be due to their large distance ($\gtrsim 300$ pc). Potentially, our analysis suggests that all disks that show a single rovibrational CO component are developing or have developed gaps in their inner 10 AU region, making them prime targets for future imaging detections.

A stellar-mass dependence of $R_{\rm CO}$ may be implied by Figure \ref{fig: final}, where higher-mass stars tend to have larger $R_{\rm CO}$ relative to lower-mass stars. CO gas from $\lesssim2$\,AU is rarely detected in disks around stars of $>1.5 M_{\odot}$. This cannot be explained by stellar magnetospheric truncation of the disk, which is expected to happen at radii $<0.1$\,AU across the stellar mass spectrum \citep{bouvier}, and an inner disk radius set by dust sublimation may be marginally consistent only in a few disks, and only at $<1$\,AU given their bolometric luminosities \citep{Dull10,vdplas15}. Alternatively, a stellar-mass dependent inner disk lifetime would be consistent with the CO temperature-radius diagram, with disks around intermediate-mass stars dispersing faster than those around low- and solar-mass stars \citep[e.g.][]{muzerolle10,yasui,ribas15}. That is, we detect fewer broad CO components in disks with short dispersal time scales. In this scenario, a mass-dependence in Figure \ref{fig: final} would be directly linked to the timescale for gas dispersal in inner disks, and therefore to planet-formation processes and timescales. 

In summary, we suggest that the CO temperature-radius diagram shows a sequence of gap opening in protoplanetary disks, where formation of smaller gaps (interpreted as an early transition, in evolutionary terms) is marked by the disappearance of the CO broad component, while opening of larger gaps (or a more evolved phase) is marked by an inversion in the trend of the temperature profile. The UV pumping regime may represent the very last step before the primordial gas in the disk is dispersed, setting the beginning of the debris disks phase \citep[see the Beta Pic point in Figure \ref{fig: final}, from rovibrational CO $v$=1-0 lines detected by][]{trout}. A high-resolution spectrum of CO emission at 4.7\,$\mu$m may therefore be a powerful discriminator between primordial, pre-transitional and transitional disk stages at radii between 0.1--10\,AU. Observations of the relative radial profiles of CO gas and of dust grains of different sizes in inner disks may provide a powerful tool to distinguish between competing gap formation mechanisms.

\subsection{Inner disk radii and the distribution of exoplanets} \label{sec: disc3}
The lifetime of the planet-forming region of protoplanetary disks is linked to the observed orbital distribution of exoplanets. In particular, the semi-major axis distribution of giant exoplanets is known to contain structure indicating the presence of multiple populations, including the ``hot Jupiters'', orbiting very close to the star at a small fraction of an AU \citep{marcy05}, and a deficit of giant planets out to 0.5\,AU \citep{cumming08}. It is generally accepted that hot Jupiters migrated inwards from a natal orbit at a few AU, either due to angular momentum exchange with protoplanetary gas \citep{lin96,kley} during the lifetime of the disk, or due to a later dynamical process, such as the Kozai mechanism \citep{kozai62, eggleton01}. 

\begin{figure}
\includegraphics[width=0.5\textwidth]{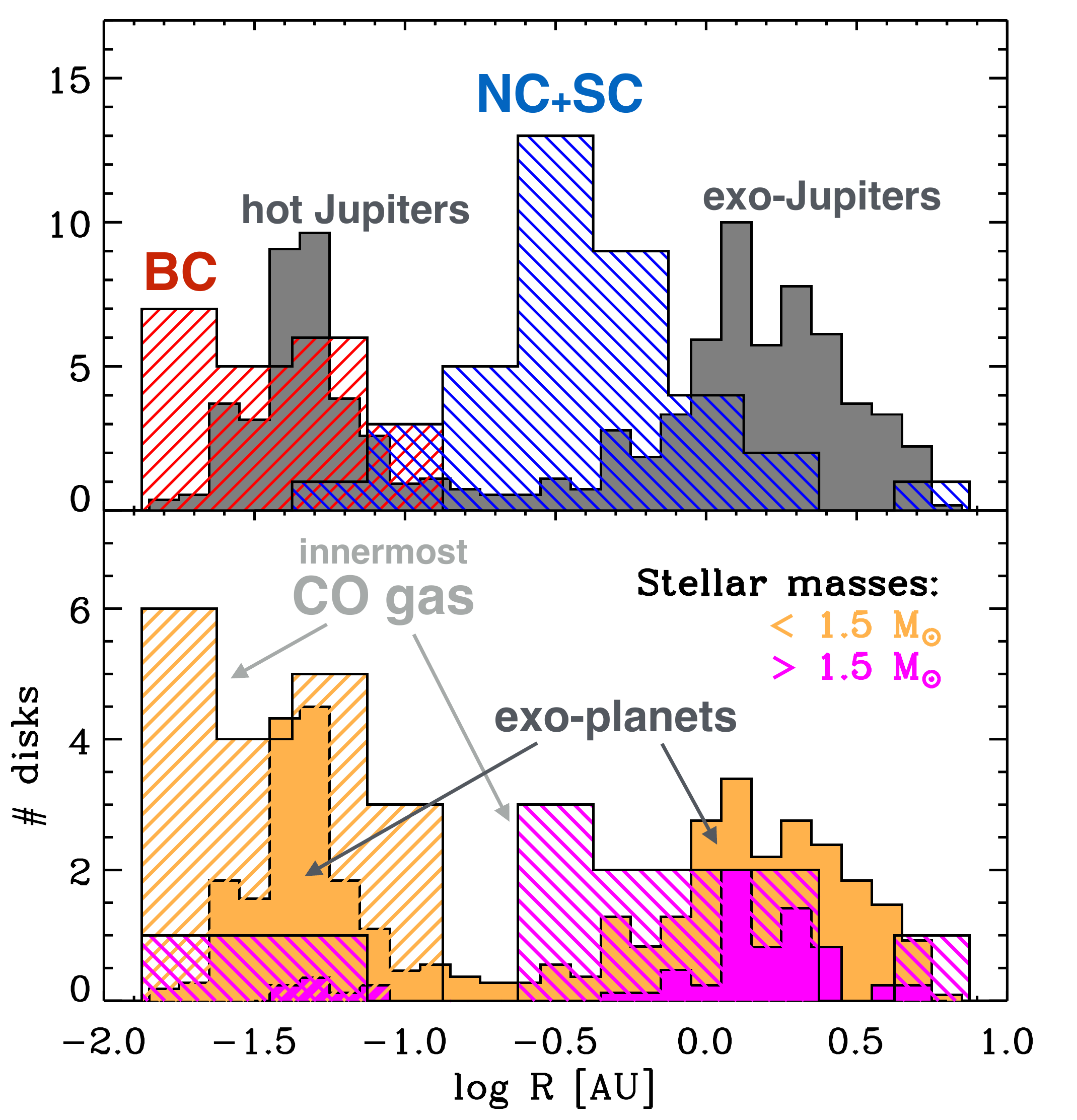} 
\caption{\textit{Top}: histograms of the inner radii of rovibrational CO emission (separated into BC and NC+SC) and of the distribution of semi-major axes of known exoplanets with M$\sin (i)>0.5 \, M_{\rm Jup}$ (from the database of exoplanets.org). The exoplanet histogram is arbitrarily scaled for comparison to the CO histogram. \textit{Bottom}: same as plot above, but separated into two stellar mass bins. Here we show the innermost CO gas in each disk, therefore excluding the NC in double-component disks.} 
\label{fig: COrad_compar}
\end{figure}

To explore this in the context of our observed inner disk radii, we show the distribution of $R_{\rm in}$ compared to the semi-major axes of known exoplanets\footnote{From {\tt exoplanets.org} \citep{exopl}; June 18 2015.} with $M\sin(i)>0.5 \, M_{\rm Jup}$ in Figure \ref{fig: COrad_compar}. The BC radius distribution matches the orbital distribution of hot Jupiters (0.01--0.1 AU), as expected in the Type II migration scenario. That is, we observe that gas is indeed often available at the right disk radii within 1--2\,Myr to allow hot Jupiters to migrate inwards to the inner gas disk truncation radius. However, hot Jupiters probably comprise a small fraction of giant exoplanets \citep{marcy05}. If the inner disk dissipates rapidly, most planets may not have time to migrate, and the orbital distribution of ``cold Jupiters'' will be determined by the inner gas radius of transitional disks, as traced by the narrow CO component. Indeed, Figure \ref{fig: COrad_compar} shows that the NC distribution peaks slightly inward of the bulk of the exo-Jupiter distribution (1--10 AU). 

A stellar-mass dependence of the inner disk gas radius may also be reflected in the exoplanet population. It is known that intermediate-mass stars have very few hot Jupiters compared to lower-mass stars \citep{John07,sato08}, although a possible overestimation of the intermediate-mass stellar masses is currently under debate \citep{Lloyd11,John13}. \cite{cur09a} proposed that the paucity of hot Jupiters around intermediate-mass stars may be due to a stellar-mass dependent disk lifetime, where inner disks around intermediate-mass stars dissipate before any giant planets have time to migrate to the disk corotation radius. In the lower panel of Figure \ref{fig: COrad_compar}, we compare the distributions of innermost gas disk radii separated into two stellar mass bins at $M_{\star} = 1.5\,M_{\odot}$. This distribution suggests that the BC component is rarely found in disks around intermediate-mass stars, relative to disks around low- and solar-mass disks. That is, the paucity of the BC within 0.1\,AU in disks around intermediate-mass stars is reflected by a similar paucity of hot Jupiters in the same stellar mass bin. A likely interpretation of our CO observations is that the gaseous disk at 0.01-0.5\,AU around stars with $M_{\star} > 1.5\,M_{\odot}$ indeed dissipates faster than the planet migration time scale. Larger samples of CO rovibrational spectra of disks and exoplanets around intermediate-mass stars are needed to confirm this link between the natal disk properties and planetary architectures.

\section{Summary \& Conclusions} \label{sec:end}
We presented a new analysis of VLT-CRIRES high-resolution (R$\sim$96,000) 4.7\,$\mu$m spectra of rovibrational CO emission from protoplanetary disks. We applied an empirical decomposition of CO line profiles to the sample of CO-emitting disks, finding that rovibrational line profiles consist of two distinct velocity components, a broad (BC) and a narrow (NC) component, primarily in disks around solar-mass stars, and of a single CO component in transitional and Herbig disks. When splitting the observed line profiles into two components, we find an empirical relation between the CO line width and the disk inclination as expected when the gas is in Keplerian rotation in a disk. This allowed us to estimate the inclination for 15 disks in the sample for which it was previously unknown.

Using the profile decomposition to separate blended information from different disk radii, we define an empirical diagram of the Keplerian CO emitting radius and the vibrational excitation temperature, leading to the definition of a universal excitation temperature profile. Between 0.05 and 3\,AU, we find a power law $T_{\rm vib. ex.} \propto R^{-0.3\pm0.1}$. Beyond 3\,AU, the vibrational temperature turns around and begins to increase with radius. We interpret this as the presence of two different excitation regimes, with pumping by infrared photons dominating in the inner disk, and fluorescence excitation by UV photons beyond 3\,AU. We discuss the CO temperature-radius diagram in the context of the theoretical picture of inside-out protoplanetary disk evolution, where the innermost disk is dissipated first. We find evidence that the CO rovibrational line profiles are tracers of the evolutionary stage of inner protoplanetary disks, potentially providing a gaseous equivalent to the traditional definition of classical, pre-transitional, and transitional disks as based on the dust continuum emission. Specifically, we propose that disks exhibiting only a single component in rovibrational CO should be in more advanced stages of gap opening and inner disk dissipation than the double-component disks, where the disappearance of BC marks an earlier stage and the UV pumping regime a final stage before the debris disks phase.

Rovibrational CO emission is sensitive to the innermost location of abundant gas, including for transitional disks, which may have more than one inner region due to the presence of gaps. Consequently, we compared the observed distribution of inner disk molecular gas to the semi-major axis distribution of giant exoplanets. We find that the distribution of CO emitting radii for the broad component matches that of the semi-major axis distribution of hot Jupiters. This is consistent with the majority of hot Jupiters migrating inward due to Type II migration, and being halted by the inner truncation of the gaseous disk during an early evolutionary phase. The radial distribution of the narrow CO component peaks at larger radii, close to the ``period deficit'' between the hot Jupiter population and the bulk of the giant exoplanet distribution. Finally, we find that disks around intermediate-mass stars rarely have gas in the hot Jupiter region. This matches an observed deficit of hot Jupiters around intermediate-mass stars, lending support to the idea that their inner disks may evolve too fast for giant planets to migrate inwards to the disk corotation radius.

This analysis ultimately supports the potential of future spatially-resolved imaging of rovibrational CO to move significant steps forward in our understanding of disk evolution and planet formation in protoplanetary disks. Herbig disks can already be resolved with adaptive-optics assisted observations, for instance with CRIRES, but disks around solar-mass stars require either spectro-astrometric techniques, infrared interferometric imaging (e.g., VLT-MATISSE), or full aperture spectro-imaging at 4.7\,$\mu$m with the future generation of extremely large telescopes (e.g., E-ELT-METIS).

\acknowledgements 
The authors acknowledge helpful discussions with C. Salyk, E. van Dishoeck, and G. Blake, and constructive comments from an anonymous referee.
This research has made use of the Exoplanet Orbit Database and the Exoplanet Data Explorer at exoplanets.org.
A.B. acknowledges financial support by a NASA Origins of the Solar System grant No. OSS 11-OSS11-0120, a NASA Planetary Geology and Geophysics Program under grant NAG 5-10201. This work is based on observations made with ESO telescopes at the Paranal Observatory under programs 179.C-0151 and 093.C-0432.

\appendix

\section{CO vibrational ratios} \label{app: vibr_convers}
\begin{figure*}
\includegraphics[width=0.5\textwidth]{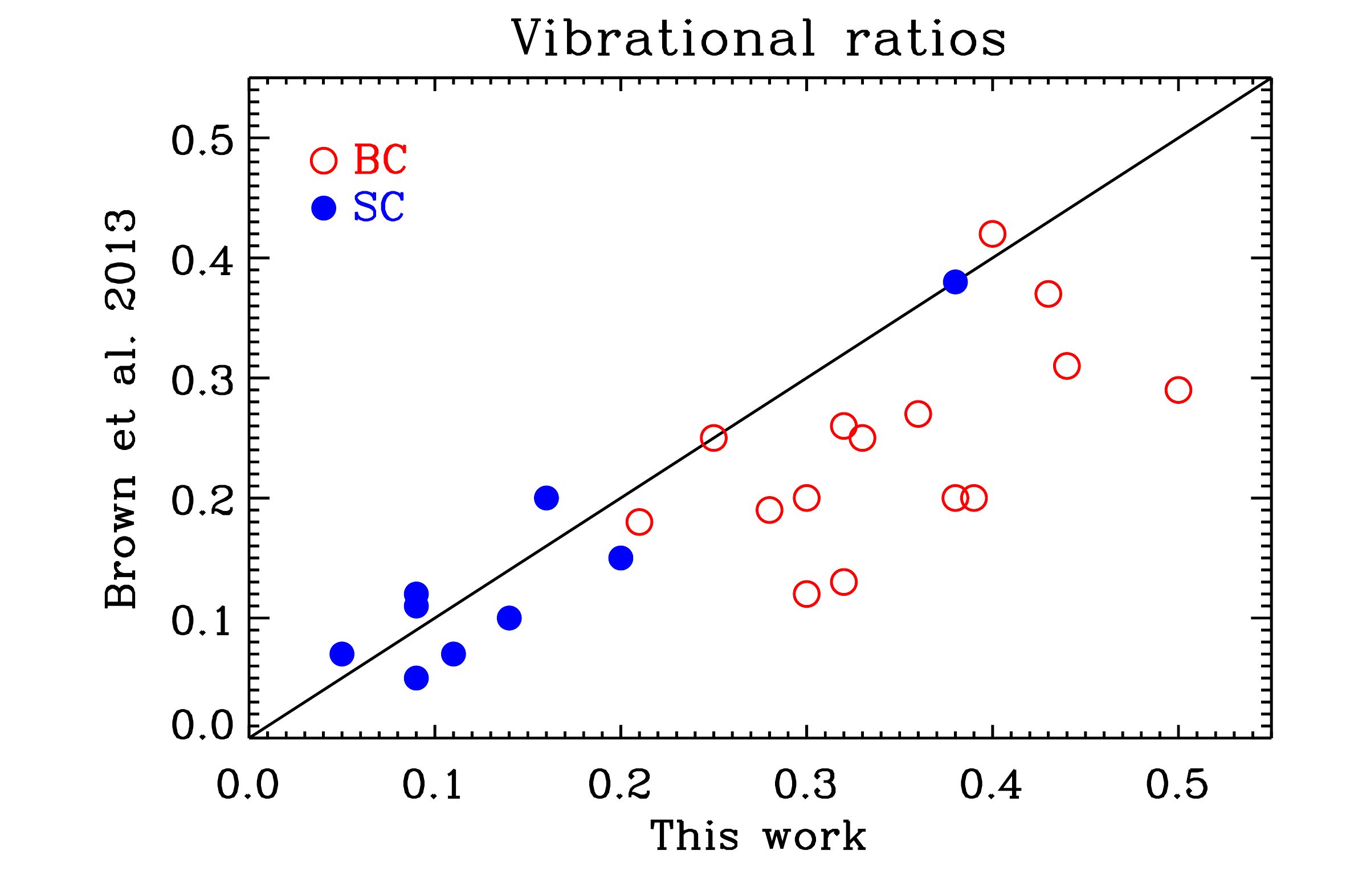} 
\includegraphics[width=0.5\textwidth]{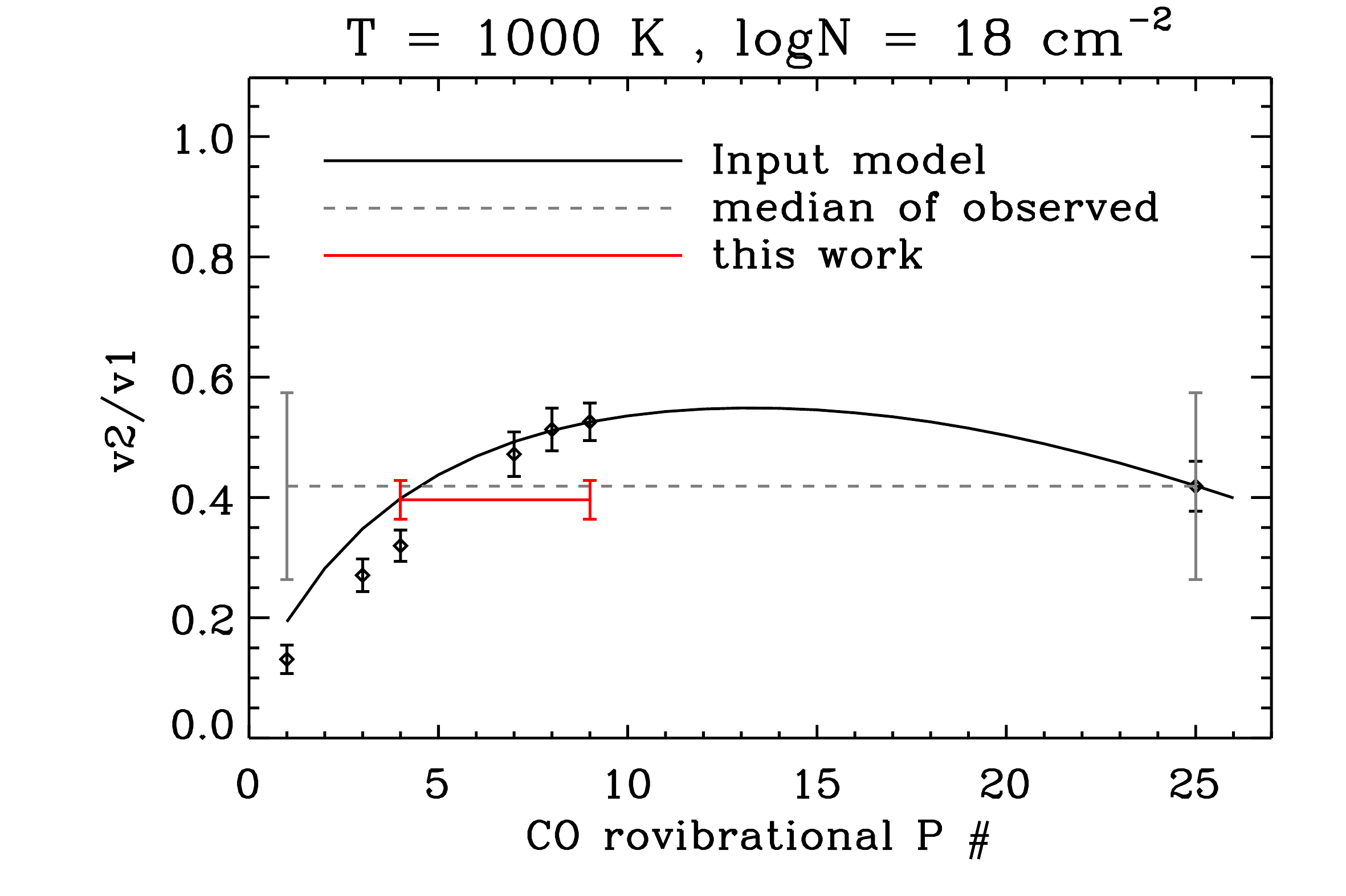} 
\includegraphics[width=0.5\textwidth]{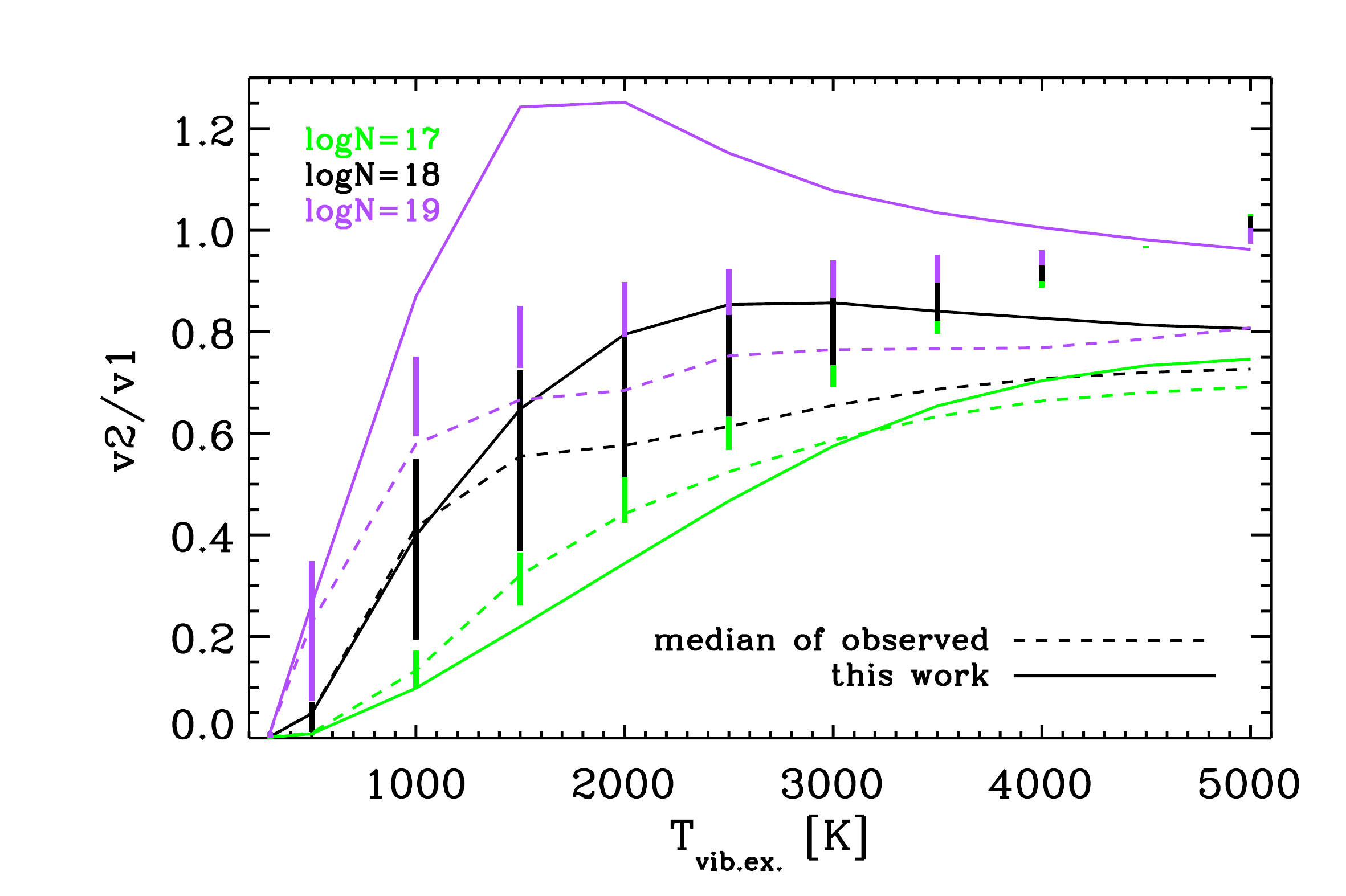} 
\includegraphics[width=0.5\textwidth]{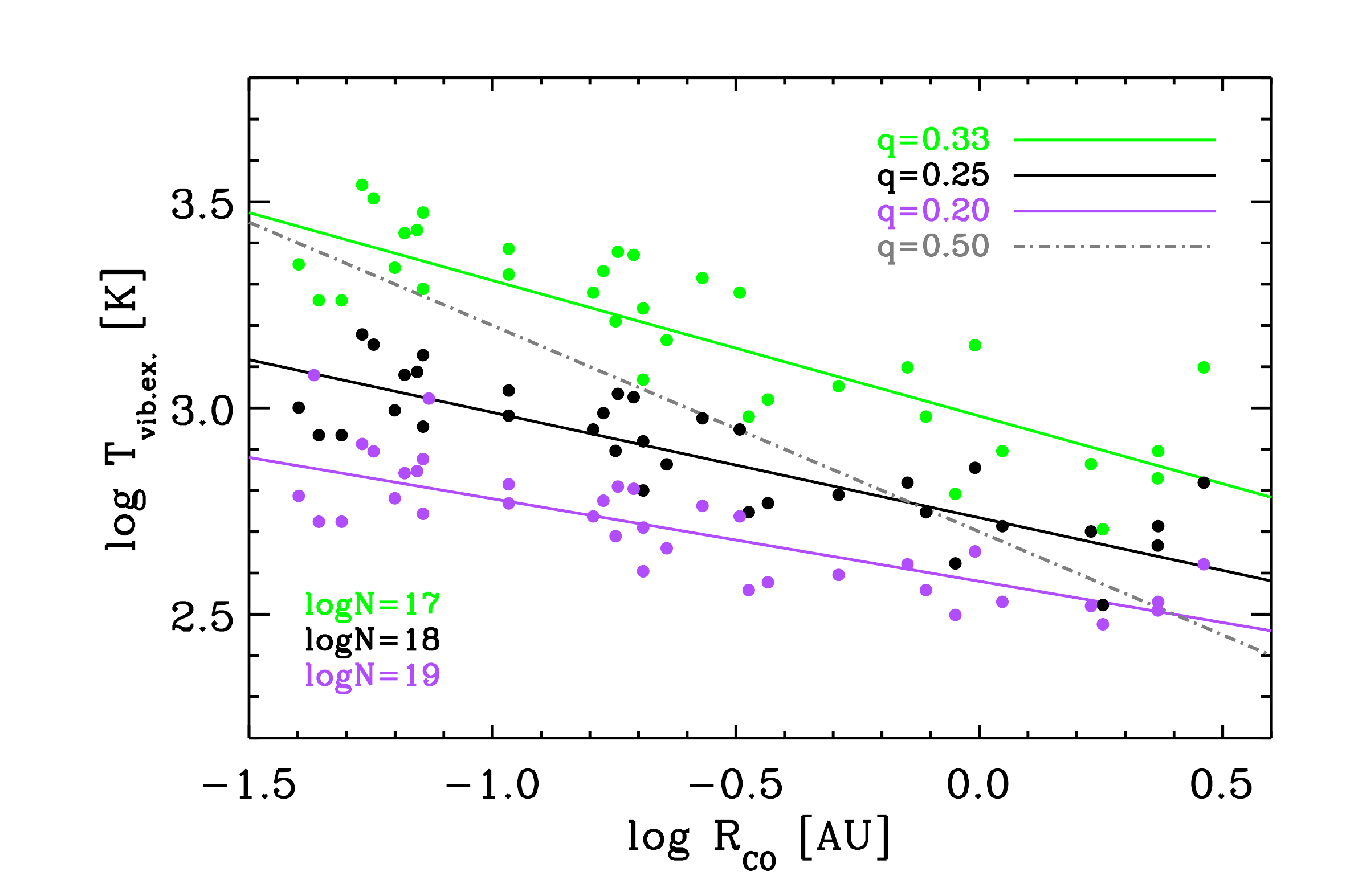} 
\caption{In order from top left to bottom right. \textit{Panel a}: comparison of vibrational ratios measured in this work to those measured in \citet{brown13}. The 1:1 relation is marked with a solid line. \textit{Panel b}: comparison of measured values for $v2/v1$ in a model with T = 1000 K and N = $10^{18}$\,cm$^{-2}$, showing the range of P-branch rovibrational levels covered by the CRIRES spectra; the true value is shown with a solid black curve, while the method adopted in \citet{brown13} and in this work are shown in grey-dashed and red-solid lines respectively. \textit{Panel c}: vibrational ratio as a function of temperature and column density over the grid of models explored; the range of true values for each model is shown as a vertical thick solid line. \textit{Panel d}: radial temperature profiles obtained after conversion of the measured $v2/v1$ values into temperature values as shown in \textit{panel c} for three values of the CO column density. An upper limit to the index $q$ is overplotted in dashed-dotted line, consistent with the data in the case of a gradient of CO column densities increasing with disk radius.}
\label{fig: vibr_Plev}
\end{figure*}

The vibrational ratio $v2/v1$ is the flux ratio $F_{v=2-1}/F_{v=1-0}$ between the $v=2-1$ lines and the $v=1-0$ lines with same rotational quantum numbers in the two vibrational levels. In this work, $v2/v1$ values are measured from stacked lines built by using multiple rotational lines and thus provide an average estimate of the vibrational ratio between $v=2$ and $v=1$. Previously, vibrational ratios of CO emission were measured by \citet{brown13} for 25 disks that are included in our sample, using the same CRIRES spectra we use here. They published the median of five values measured in each disk, as obtained from ratios of the corresponding rotational lines in the two vibrational levels. Figure \ref{fig: vibr_Plev} (panel $a$) compares the values from \citet{brown13} to those measured in this work. The values measured in the single-component disks agree to a 1:1 relation, showing that our estimates are consistent. However, the vibrational ratios in previous work have been underestimated in the double-component disks, because the different contributions from BC and NC in the $v$=1-0 lines were not separated.

The vibrational ratio $v2/v1$ depends on the temperature and column density of the gas, as well as on the excitation process. In this work we use a model of a slab of gas in thermal equilibrium that accounts for the line opacity, where the level populations are determined by the Boltzmann distribution \citep[the model is explicitly described in][]{banz12}. The molecular parameters for CO rovibrational lines are taken from the 2012 HITRAN and the HITEMP databases \citep{hitemp,hitran12}. We produced a grid of models to convert the measured $v2/v1$ into an estimate of the vibrational excitation temperature $T_{\rm vib. ex.}$ of the gas. We used three values for the CO column density N$_{\rm{co}}$ to span the range of values proposed by previous studies of NIR CO emission from protoplanetary disks \citep[$10^{17}$--$10^{19}$\,cm$^{-2}$,][]{blake04,greg,sal11,banz15,vdplas15}. Similarly, for the temperature we use eleven values between 300 and 5000 K. For each set of $T_{\rm vib. ex.}$ and $N_{\rm CO}$ a CO model spectrum is produced, assuming a line FWHM of $\sim75$\,km\,s$^{-1}$ (the median value for the BC) and using a pixel sampling equivalent to the CRIRES spectra. We then re-sample the model spectrum using a typical value for the rms noise as measured in the observed spectra, and then measure vibrational ratios following three methods: 1) using directly the individual line fluxes before convolution with the observed FWHM (the ``true" value), 2) taking the median of seven ratios of the line fluxes measured in the convolved spectrum \citep[equivalent to the method used by][]{brown13}, and 3) using the method adopted in this work. Figure \ref{fig: vibr_Plev} (panel $b$) shows one model as an example, for illustration of the outcomes of this procedure. It is important to highlight that the ``true" ratio is not a single value. In the example, the vibrational ratio varies between 0.2 and 0.6 depending on the line used to measure it. Moreover, vibrational ratios measured from the individual line fluxes as observed in the convolved spectrum tend to underestimate the true value. This is due to blending of the $v$=1-0 lines with higher vibrational lines, which contribute increasingly to the line flux as $T_{\rm vib. ex.}$, $N_{\rm CO}$, and the line FWHM increase. The averaged ratios estimated in this work in general agree well with the range of true values for each model (panel $c$ in Figure \ref{fig: vibr_Plev}). This is true in particular in the case of a CO column density of $10^{18}$\,cm$^{-2}$, which is a good approximation for most disks in our sample \citep[e.g.][]{sal11}. For higher values of $N_{\rm CO} = 10^{19}$\,cm$^{-2}$, the vibrational ratios measured here may overestimate the true values by up to 50\%.

We convert the vibrational ratios measured in this work into temperatures using the curves shown in panel $c$ of Figure \ref{fig: vibr_Plev}. In this regard, it is not important that our measured $v2/v1$ may overestimate the true value, as long as we are self-consistent in the conversion, i.e. as long as we use the same method to measure $v2/v1$ in the data and in the model. In Tables \ref{tab:2co comp} and \ref{tab:1co comp} we put the temperature values estimated by using a CO column of $10^{18}$\,cm$^{-2}$. In the case of AS 209, HH 100, and IM Lup, the measured vibrational ratios are too high to be modeled with a CO column of $10^{18}$\,cm$^{-2}$, which never produces values higher than $\sim$0.8. Therefore, for these disks we use a CO column of $10^{19}$\,cm$^{-2}$, which can reproduce $v2/v1$ as high as 1.2 and would be appropriate if the emission is more optically thick (our analysis indeed finds that the BC in AS 209 and IM Lup is more optically thick than in other double-component disks, as the BC is detected also in $^{13}$CO, see Figure \ref{fig: multi-comp analysis}). For each one of the three values adopted for the CO column density, we obtain a different empirical relation between the temperature and the radius of the CO emission. This is shown in panel $d$ of Figure \ref{fig: vibr_Plev}, where we plot the linear fits obtained as in Figure \ref{fig: vibr_rin}. The index $q$ of the temperature profile $T_{\rm vib. ex.} \propto R^{-q}$ varies between 0.20 (for $N_{\rm CO}$ = $10^{19}$\,cm$^{-2}$) to 0.33 (for $N_{\rm CO} = 10^{17}$\,cm$^{-2}$). If we allow a radial gradient in $N_{\rm CO}$ such to have higher values of $\sim 10^{18}-10^{19}$\,cm$^{-2}$ as representative of the more optically-thick NC, and lower values of $\sim 10^{17}-10^{18}$\,cm$^{-2}$ as representative of the BC \citep[as found by][]{greg}, the index may become as steep as $\sim 0.5$.

\section{CO decomposition of the entire disk sample} \label{app: CO_decomp}
\begin{figure*}
\includegraphics[width=1\textwidth]{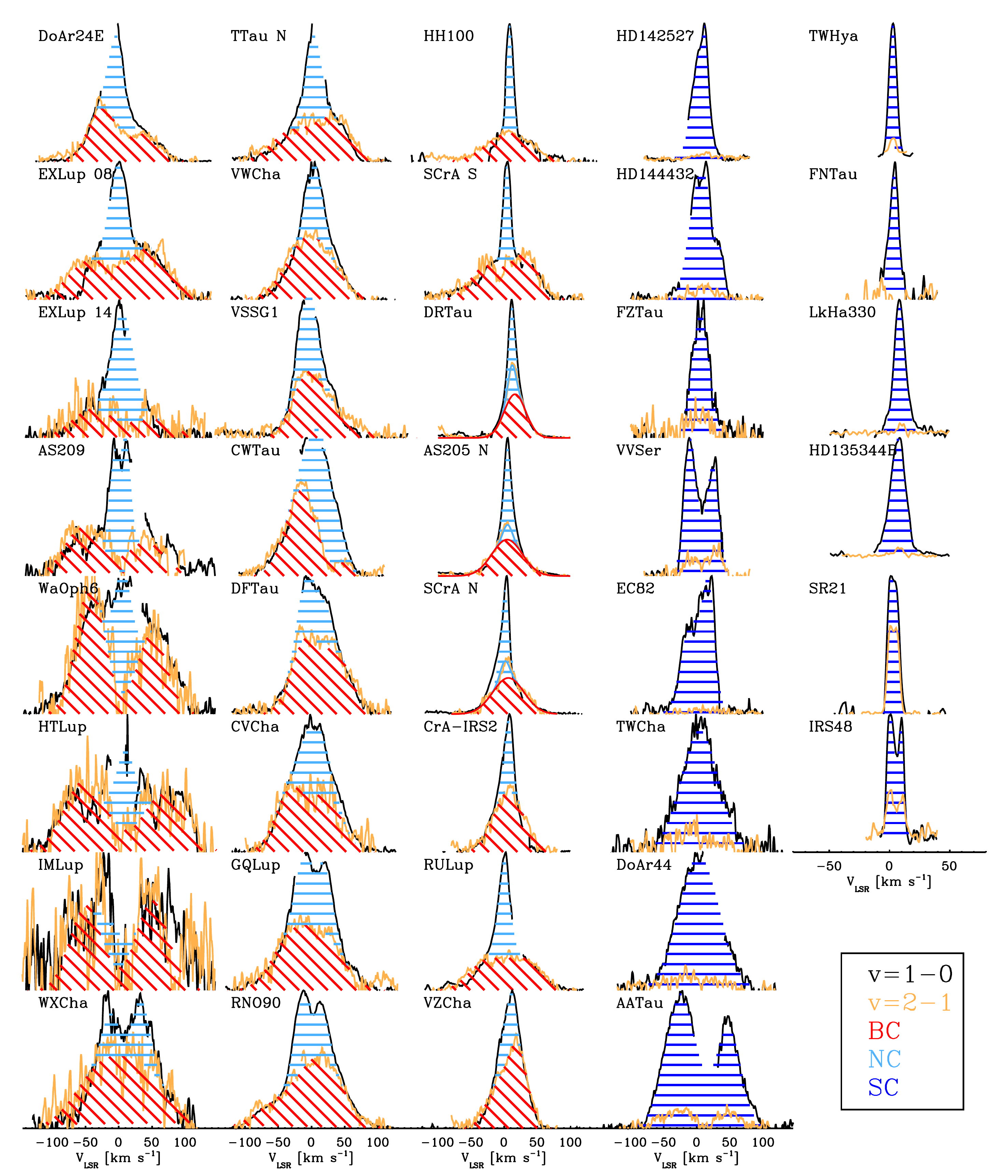} 
\caption{CO decomposition plots of the entire disk sample. The color-coding follows what used in Figure \ref{fig: multi-comp analysis}: $^{12}$CO $v=1-0$ lines are shown in black, $^{12}$CO $v=2-1$ lines in orange; NC is filled with cyan horizontal lines, BC with red diagonal lines, SC with blue horizontal lines. In the double-component disks (first three columns from the left), $^{12}$CO $v=2-1$ lines are scaled up to the $^{12}$CO $v=1-0$ lines to illustrate the two CO components. In the three disks where NC is detected also in the $^{12}$CO $v=2-1$ - DRTau, AS205 N, and SCrA N - Gaussian fits to BC and NC are shown on top of the line. Single-component disks are shown in the last two columns to the right.}
\label{fig: decomp figure}
\end{figure*}

In Figure \ref{fig: decomp figure} we show the outcome of the CO decomposition procedure as applied to all disk spectra in our sample. 

\section{Bayesian method for linear regressions} \label{app: bayes}
\begin{figure*}
\includegraphics[width=1\textwidth]{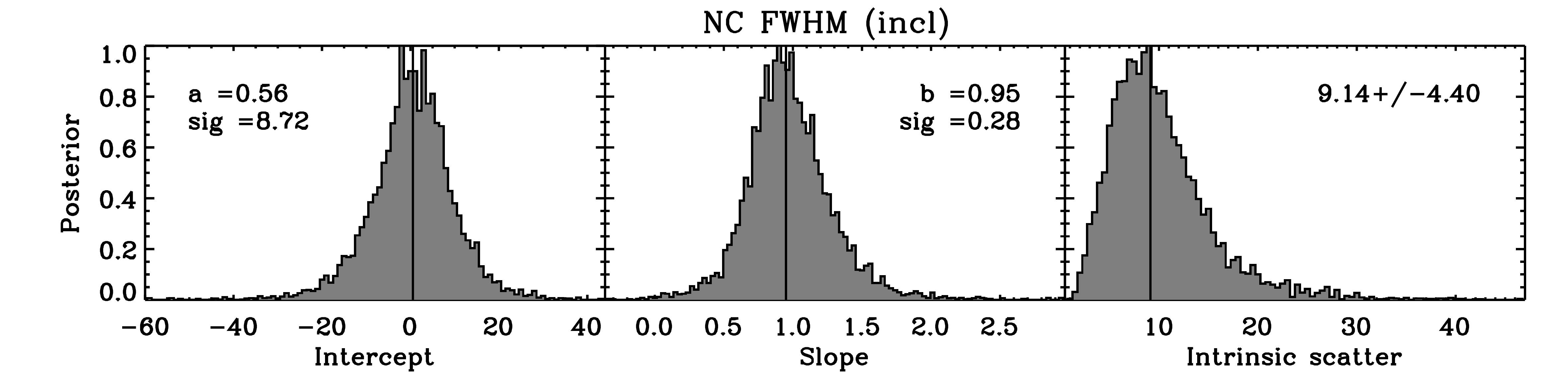} 
\includegraphics[width=1\textwidth]{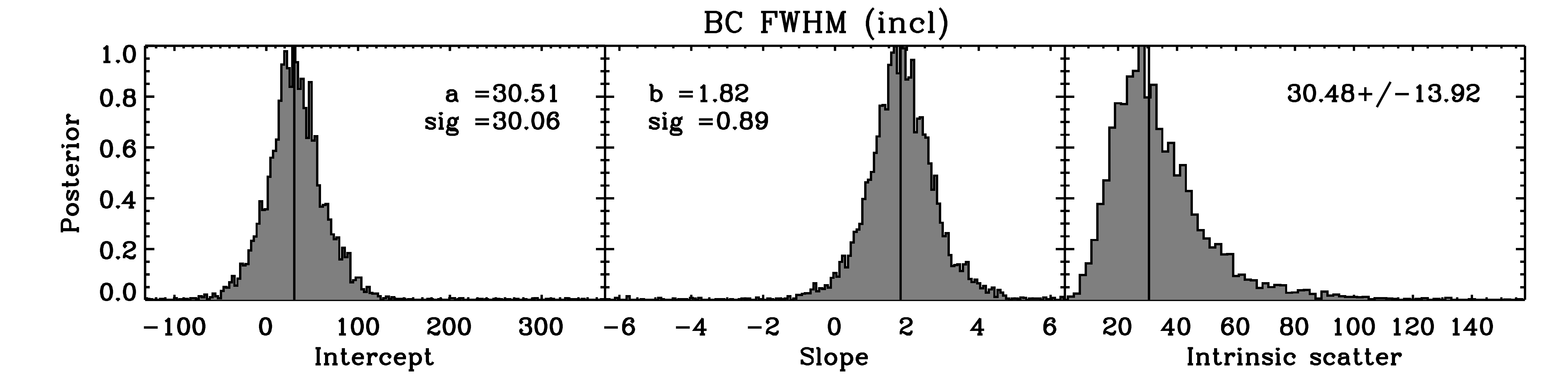} 
\includegraphics[width=1\textwidth]{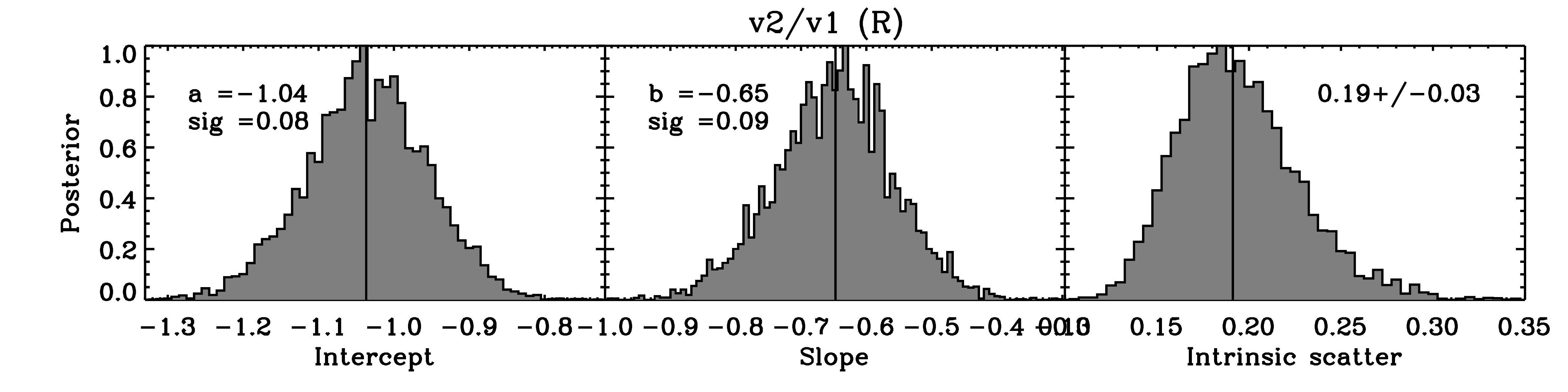} 
\includegraphics[width=1\textwidth]{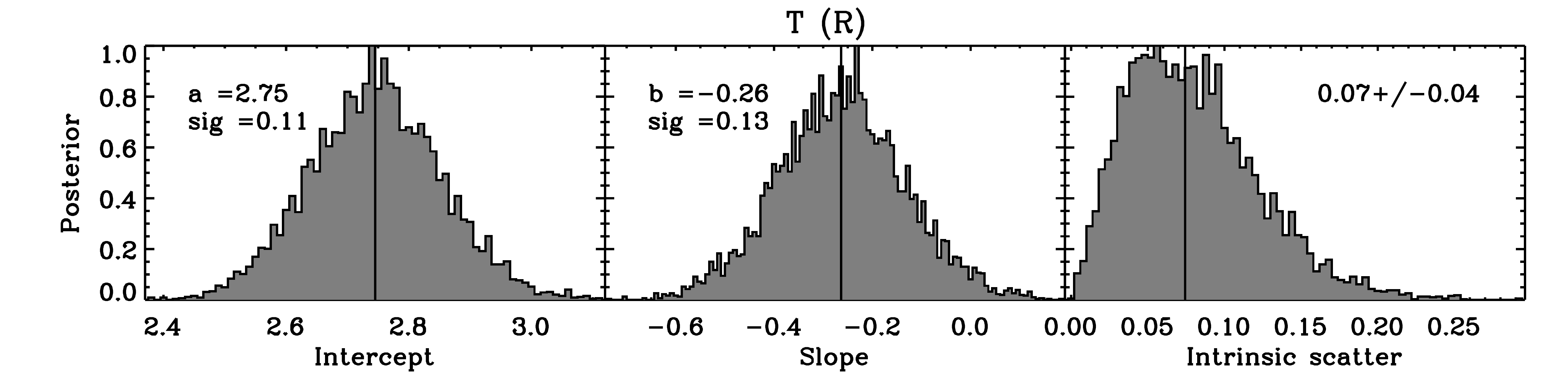} 
\caption{Posterior distributions for linear regression parameters obtained with the method by \citet{kelly} for the empirical relations presented in Section \ref{sec:ana}. The first two plots from the top are for the empirical relations between the BC and NC FWHM and the disk inclination (Figure \ref{fig: fwhm_incl}), while the last two plots are for the relations between vibrational ratios, or temperature, and disk radius (Figure \ref{fig: vibr_rin}). Vertical lines mark the median value of each distribution, which is reported inside each plot together with a robust estimate of the posterior standard deviation.}
\label{fig: post_distr}
\end{figure*}

We implement the Bayesian method by \citet{kelly} in the linear fits to the empirical relations presented in Sections \ref{sec:ana3} and \ref{sec:ana4}. \citet{kelly} provided a Bayesian method for linear regression that accounts for errors on both the dependent and independent variables as well as an intrinsic scatter due to physical properties that are not explicitly included. A weighted sum of Gaussian distributions is used to obtain an accurate approximation of the true probability distribution of the independent variable. To sample the posterior distribution we use the Markov chain Monte Carlo method from the Metropolis-Hastings algorithm, which is more efficient than the Gibbs sampler in case of small sample sizes. A thorough discussion of the method and of its comparison to other statistical methods is provided in \citet{kelly}. The method has been made publicly available through the IDL routine \texttt{linmix\_err.pro}, that we use in our analysis. Figure \ref{fig: post_distr} shows the posterior distributions we obtained for the linear regression parameters presented in Section \ref{sec:ana}. 

\newpage

\begin{thebibliography}{}

\bibitem[Alexander et al.(2006)]{alex06} Alexander, R.~D., Clarke, C.~J., \& Pringle, J.~E.\ 2006, \mnras, 369, 216 

\bibitem[Alexander et al.(2014)]{alex14} Alexander, R., Pascucci, I., Andrews, S., Armitage, P., \& Cieza, L.\ 2014, Protostars and Planets VI, 475 

\bibitem[Andrews et al.(2009)]{and09} Andrews, S.~M., Wilner, D.~J., Hughes, A.~M., Qi, C., \& Dullemond, C.~P.\ 2009, \apj, 700, 1502 

\bibitem[Armitage(2011)]{armi} Armitage, P.~J.\ 2011, \araa, 49, 195 

\bibitem[Avenhaus et al.(2014)]{Avenhaus} Avenhaus, H., Quanz, S.~P., Schmid, H.~M., et al.\ 2014, \apj, 781, 87 

\bibitem[Banzatti et al.(2012)]{banz12} Banzatti, A., Meyer, M.~R., Bruderer, S., et al.\ 2012, \apj, 745, 90 

\bibitem[Banzatti et al.(2015)]{banz15} Banzatti, A., Pontoppidan, K.~M., Bruderer, S., Muzerolle, J., \& Meyer, M.~R.\ 2015, \apjl, 798, LL16

\bibitem[Bast et al.(2011)]{bast} Bast, J.~E., Brown, J.~M., Herczeg, G.~J., van Dishoeck, E.~F., \& Pontoppidan, K.~M.\ 2011, \aap, 527, AA119 

\bibitem[Beckwith et al.(1990)]{beck} Beckwith, S.~V.~W., Sargent, A.~I., Chini, R.~S., \& Guesten, R.\ 1990, \aj, 99, 924 

\bibitem[Blake \& Boogert(2004)]{blake04} Blake, G.~A., \& Boogert, A.~C.~A.\ 2004, \apjl, 606, L73 

\bibitem[Bouvier et al.(1999)]{bouv99} Bouvier, J., Chelli, A., Allain, S., et al.\ 1999, \aap, 349, 619 

\bibitem[Bouvier et al.(2007)]{bouvier} Bouvier, J., Alencar, S.~H.~P., Harries, T.~J., Johns-Krull, C.~M., \& Romanova, M.~M.\ 2007, Protostars and Planets V, 479 

\bibitem[Brittain et al.(2003)]{brit03} Brittain, S.~D., Rettig, T.~W., Simon, T., et al.\ 2003, \apj, 588, 535 

\bibitem[Brittain et al.(2007)]{brit07} Brittain, S.~D., Simon, T., Najita, J.~R., \& Rettig, T.~W.\ 2007, \apj, 659, 685 

\bibitem[Brown et al.(2009)]{brown09} Brown, J.~M., Blake, G.~A., Qi, C., et al.\ 2009, \apj, 704, 496 

\bibitem[Brown et al.(2012)]{brown12} Brown, J.~M., Herczeg, G.~J., Pontoppidan, K.~M., \& van Dishoeck, E.~F.\ 2012, \apj, 744, 116 

\bibitem[Brown et al.(2013)]{brown13} Brown, J.~M., Pontoppidan, K.~M., van Dishoeck, E.~F., et al.\ 2013, \apj, 770, 94 

\bibitem[Calvet et al.(2002)]{calvet} Calvet, N., D'Alessio, P., Hartmann, L., et al.\ 2002, \apj, 568, 1008 

\bibitem[Coffey et al.(2008)]{coffey} Coffey, D., Bacciotti, F., \& Podio, L.\ 2008, \apj, 689, 1112 

\bibitem[Cox et al.(2013)]{cox} Cox, A.~W., Grady, C.~A., Hammel, H.~B., et al.\ 2013, \apj, 762, 40 

\bibitem[Cumming et al.(2008)]{cumming08} Cumming, A., Butler, R.~P., Marcy, G.~W., et al.\ 2008, \pasp, 120, 531 

\bibitem[Currie(2009)]{cur09a} Currie, T.\ 2009, \apjl, 694, L171

\bibitem[Dullemond \& Monnier(2010)]{Dull10} Dullemond, C.~P., \& Monnier, J.~D.\ 2010, \araa, 48, 205 

\bibitem[Eggleton \& Kiseleva-Eggleton(2001)]{eggleton01} Eggleton, P.~P., \& Kiseleva-Eggleton, L.\ 2001, \apj, 562, 1012 

\bibitem[Espaillat et al.(2007)]{espaillat07} Espaillat, C., Calvet, N., D'Alessio, P., et al.\ 2007, \apjl, 670, L135 

\bibitem[Feigelson et al.(1993)]{feig} Feigelson, E.~D., Casanova, S., Montmerle, T., \& Guibert, J.\ 1993, \apj, 416, 623 

\bibitem[Geers et al.(2007)]{geers07} Geers, V.~C., Pontoppidan, K.~M., van Dishoeck, E.~F., et al.\ 2007, \aap, 469, L35 

\bibitem[Gorti et al.(2015)]{gorti15} Gorti, U., Hollenbach, 
D., \& Dullemond, C.~P.\ 2015, \apj, 804, 29 

\bibitem[Goto et al.(2011)]{goto} Goto, M., Reg{\'a}ly, Z., Dullemond, C.~P., et al.\ 2011, \apj, 728, 5 

\bibitem[Gras-Vel{\'a}zquez \& Ray(2005)]{grasvel} Gras-Vel{\'a}zquez, {\`A}., \& Ray, T.~P.\ 2005, \aap, 443, 541 

\bibitem[Han et al.(2014)]{exopl} Han, E., Wang, S.~X., Wright, J.~T., et al.\ 2014, \pasp, 126, 827 

\bibitem[Herczeg et al.(2011)]{greg} Herczeg, G.~J., Brown, J.~M., van Dishoeck, E.~F., \& Pontoppidan, K.~M.\ 2011, \aap, 533, A112 

\bibitem[Isella et al.(2009)]{isella09} Isella, A., Carpenter, J.~M., \& Sargent, A.~I.\ 2009, \apj, 701, 260 

\bibitem[Kaeufl et al.(2004)]{crires} Kaeufl, H.-U., Ballester, P., Biereichel, P., et al.\ 2004, \procspie, 5492, 1218

\bibitem[Kelly(2007)]{kelly} Kelly, B.~C.\ 2007, \apj, 665, 1489 

\bibitem[Kley \& Nelson(2012)]{kley} Kley, W., \& Nelson, R.~P.\ 2012, \araa, 50, 211 

\bibitem[Kozai(1962)]{kozai62} Kozai, Y.\ 1962, \aj, 67, 591 

\bibitem[Krotkov et al.(1980)]{krotkov} Krotkov, R., Wang, D., 
\& Scoville, N.~Z.\ 1980, \apj, 240, 940 

\bibitem[Kudo et al.(2008)]{kudo} Kudo, T., Tamura, M., Kitamura, Y., et al.\ 2008, \apjl, 673, L67 

\bibitem[Johnson et al.(2007)]{John07} Johnson, J.~A., Fischer, D.~A., Marcy, G.~W., et al.\ 2007, \apj, 665, 785 

\bibitem[Johnson et al.(2013)]{John13} Johnson, J.~A., Morton, 
T.~D., \& Wright, J.~T.\ 2013, \apj, 763, 53 

\bibitem[Lin et al.(1996)]{lin96} Lin, D.~N.~C., Bodenheimer, P., \& Richardson, D.~C.\ 1996, \nat, 380, 606 

\bibitem[Lloyd(2011)]{Lloyd11} Lloyd, J.~P.\ 2011, \apjl, 739, 
L49

\bibitem[Luhman et al.(2010)]{luhman10} Luhman, K.~L., Allen, P.~R., Espaillat, C., Hartmann, L., \& Calvet, N.\ 2010, \apjs, 186, 111 

\bibitem[Marcy et al.(2005)]{marcy05} Marcy, G., Butler, R.~P., Fischer, D., et al.\ 2005, Progress of Theoretical Physics Supplement, 158, 24 

\bibitem[Maaskant et al.(2013)]{maask13} Maaskant, K.~M., Honda, M., Waters, L.~B.~F.~M., et al.\ 2013, \aap, 555, A64 

\bibitem[McClure et al.(2013)]{mccl13} McClure, M.~K., Calvet, N., Espaillat, C., et al.\ 2013, \apj, 769, 73 

\bibitem[Muzerolle et al.(2010)]{muzerolle10} Muzerolle, J., Allen, L.~E., Megeath, S.~T., Hern{\'a}ndez, J., \& Gutermuth, R.~A.\ 2010, \apj, 708, 1107 

\bibitem[Najita et al.(2003)]{naji03} Najita, J., Carr, J.~S., \& Mathieu, R.~D.\ 2003, \apj, 589, 931 

\bibitem[P{\'e}rez et al.(2014)]{perez14} P{\'e}rez, L.~M., Isella, A., Carpenter, J.~M., \& Chandler, C.~J.\ 2014, \apjl, 783, L13 

\bibitem[Pinilla et al.(2015)]{pinilla15} Pinilla, P., de Juan Ovelar, M., Ataiee, S., Benisty, M., Birnstiel, T., van Dishoeck, E.~F., \& Min, M.\ 2015, \aap, 573, A9 

\bibitem[Pinte et al.(2008)]{pinte} Pinte, C., Padgett, D.~L., M{\'e}nard, F., et al.\ 2008, \aap, 489, 633 

\bibitem[Pontoppidan et al.(2007)]{pont07} Pontoppidan, K.~M., Dullemond, C.~P., Blake, G.~A., et al.\ 2007, \apj, 656, 980 

\bibitem[Pontoppidan et al.(2008)]{pont08} Pontoppidan, K.~M., Blake, G.~A., van Dishoeck, E.~F., et al.\ 2008, \apj, 684, 1323 

\bibitem[Pontoppidan et al.(2011b)]{pont11} Pontoppidan, K.~M., Blake, G.~A., \& Smette, A.\ 2011a, \apj, 733, 84 

\bibitem[Pontoppidan et al.(2011a)]{pont_msgr} Pontoppidan, K.~M., van Dishoeck, E., Blake, G.~A., et al.\ 2011b, The Messenger, 143, 32 

\bibitem[Ratzka et al.(2009)]{ratzka} Ratzka, T., Schegerer, A.~A., Leinert, C., et al.\ 2009, \aap, 502, 623 

\bibitem[Ribas et al.(2015)]{ribas15} Ribas, {\'A}., Bouy, H., \& Mer{\'{\i}}n, B.\ 2015, \aap, 576, A52 

\bibitem[Rigliaco et al.(2013)]{rigl13} Rigliaco, E., Pascucci, I., Gorti, U., Edwards, S., \& Hollenbach, D.\ 2013, \apj, 772, 60 

\bibitem[Rothman et al.(2010)]{hitemp} Rothman, L.~S., Gordon, I.~E., Barber, R.~J., et al.\ 2010, \jqsrt, 111, 2139 

\bibitem[Rothman et al.(2013)]{hitran12} Rothman, L.~S., Gordon, I.~E., Babikov, Y., et al.\ 2013, \jqsrt, 130, 4 

\bibitem[Salyk et al.(2009)]{sal09} Salyk, C., Blake, G.~A., Boogert, A.~C.~A., \& Brown, J.~M.\ 2009, \apj, 699, 330 

\bibitem[Salyk et al.(2011)]{sal11} Salyk, C., Blake, G.~A., Boogert, A.~C.~A., \& Brown, J.~M.\ 2011, \apj, 743, 112 

\bibitem[Salyk et al.(2013)]{sal13} Salyk, C., Herczeg, G.~J., Brown, J.~M., et al.\ 2013, \apj, 769, 21 

\bibitem[Salyk et al.(2014)]{sal14} Salyk, C., Pontoppidan, K., Corder, S., et al.\ 2014, \apj, 792, 68 

\bibitem[Sato et al.(2008)]{sato08} Sato, B., Izumiura, H., Toyota, E., et al.\ 2008, \pasj, 60, 539 

\bibitem[Thi et al.(2013)]{Thi13} Thi, W.~F., Kamp, I., Woitke, P., et al.\ 2013, \aap, 551, A49 

\bibitem[Troutman et al.(2011)]{trout} Troutman, M.~R., Hinkle, K.~H., Najita, J.~R., Rettig, T.~W., \& Brittain, S.~D.\ 2011, \apj, 738, 12 

\bibitem[van der Plas et al.(2015)]{vdplas15} van der Plas, G., van den Ancker, M.~E., Waters, L.~B.~F.~M., \& Dominik, C.\ 2015, \aap, 574, A75 

\bibitem[Unruh et al.(1998)]{Unruh} Unruh, Y.~C., Collier Cameron, A., \& Guenther, E.\ 1998, \mnras, 295, 781 

\bibitem[Yasui et al.(2014)]{yasui} Yasui, C., Kobayashi, N., Tokunaga, A.~T., \& Saito, M.\ 2014, \mnras, 442, 2543 

\bibitem[Zhu et al.(2011)]{zhu11} Zhu, Z., Nelson, R.~P., Hartmann, L., Espaillat, C., \& Calvet, N.\ 2011, \apj, 729, 47 


\end{thebibliography}
\end{document}